\documentclass[journal]{IEEEtran}
\usepackage{algorithmicx}
\usepackage{algpseudocode,algorithm2e}
\usepackage{amsmath}
\usepackage{amsfonts}
\usepackage{stfloats}

\usepackage{hyperref}
 
\usepackage{enumitem}
\usepackage{algpseudocode}
\usepackage{multirow}
\usepackage{url}
\usepackage{epsfig}
\usepackage{multicol,lipsum,xparse}
\usepackage{subcaption}
\usepackage{amsmath}
\usepackage{stmaryrd}
\usepackage{color}
\usepackage{todonotes}
\usepackage[utf8]{inputenc}
\usepackage{tikz}
\usetikzlibrary{decorations.pathreplacing}
\usepackage{graphicx}
\usepackage{colortbl}
\usepackage{tabulary}
\usepackage{etoolbox}

\usepackage{tikz}

\title{FlowTune: End-to-end Automatic Logic Optimization Exploration via Domain-specific Multi-armed Bandit}
\vspace{4mm}

\author{Walter Lau Neto \textit{IEEE Student Member}, Yingjie Li \textit{IEEE Student Member}, \\Pierre-Emmanuel Gaillardon \textit{IEEE Senior Member}, Cunxi Yu \textit{IEEE Member}}
\newbool{journalBasedAuthors}
\ifCLASSOPTIONpeerreview \setbool{journalBasedAuthors}{true}
\else \ifCLASSOPTIONjournal \setbool{journalBasedAuthors}{true}
\else \setbool{journalBasedAuthors}{false}
\fi
\fi

\ifbool{journalBasedAuthors}
{

    \author{Walter~Lau Neto, 
            Yingjie~Li,
            Pierre-Emmanuel~Gaillardon, 
            Cunxi~Yu
    \thanks{W. Lau Neto is with Silicon Realization Group of Synopsys Inc., Sunnyvale, CA, US. (e-mail: launeto@synopsys.com).}
    \thanks{Y. Li, P-E. Gaillardon, and C. Yu are with the Department of Electrical and Computer Engineering, University of Utah, Salt Lake City, US (e-mails: yingjie.li@utah.edu, pierre-emmanuel.gaillardon@utah.edu, cunxi.yu@utah.edu).}
    \thanks{This work is funded by National Science Foundation (NSF) NSF-2008144, DARPA IDEA, and NSF CAREER awards NSF-1751064 and NSF-2047176.}
    \ifCLASSOPTIONjournal
        \thanks{Digital Object Identifier 10.1109/TCAD.XXXXXXX}
    \fi
    }
}

\begin{document}

\newcommand{\pgftextcircled}[1]{
    \setbox0=\hbox{#1}%
    \dimen0\wd0%
    \divide\dimen0 by 2%
    \begin{tikzpicture}[baseline=(a.base)]%
        \useasboundingbox (-\the\dimen0,0pt) rectangle (\the\dimen0,1pt);
        \node[circle,draw,outer sep=0pt,inner sep=0.1ex] (a) {#1};
    \end{tikzpicture}
}

\maketitle

\begin{abstract}

Design flows are the explicit combinations of design transformations, primarily involved in synthesis, placement and routing processes, to accomplish the design of Integrated Circuits (ICs) and System-on-Chip (SoC). Mostly, the flows are developed based on the knowledge of the experts. However, due to the large search space of design flows and the increasing design complexity, developing \textit{Intellectual Property (IP)-specific} synthesis flows providing high Quality of Result (QoR) is extremely challenging. 

In recent years, machine learning (ML) has been increasingly used in electronic design automation (EDA), with the goal of reducing manual labor and speeding up the design closure process in current toolflows. Existing techniques, on the other hand, either necessitate a huge amount of labeled data and time-consuming training, or are constrained in terms of practical EDA toolflow integration due to computation overhead. This paper presents a generic end-to-end sequential decision making framework \textit{FlowTune} for synthesis tooflow optimization, with a novel high-performance domain-specific, multi-stage multi-armed bandit (MAB) approach. This framework addresses wide range of optimization problems on Boolean optimization problems such as And-Inv-Graphs, Conjunction Normal Form (CNF) minimization (\# clauses) for Boolean Satisfiability; logic synthesis and technology mapping, and more importantly end-to-end post place-and-route (PnR) optimizations. Moreover, we demonstrate the high extnsibility and generalizability of the proposed domain-specific MAB approach with end-to-end FPGA design flow, evaluated at post-routing stage, with two different FPGA backend tools (OpenFPGA and VPR) and two different logic synthesis representations (AIGs and MIGs). FlowTune is fully integrated with ABC, Yosys, VTR, LSOracle, OpenFPGA, and industrial tools, and is released publicly. The experimental results conducted on various design stages in the flow all demonstrate that our framework outperforms both hand-crafted flows and ML explored flows in quality of results, and is orders of magnitude faster compared to ML-based approaches.



\end{abstract}

\section{Introduction}

To manage the complicated duties needed in the design of modern Integrated Circuits (ICs), Electronic Design Automation (EDA) is made up of several phases, each relying on distinct abstract layers. Designing an effective EDA flow is a difficult and critical undertaking, given the wide range of algorithms and tuning choices available. Indeed, while EDA vendors provide generic reference design flows, a well-designed flow that is design-aware can greatly improve the \textit{Quality of Results} (QoR), as well as reduce the time-to-market by reducing the number of iterations to achieve design closure. Existing CAD tools do not ensure design closure in an out-of-the-box approach, which is a major roadblock to rapid hardware specialization. To achieve a high QoR, these tools normally necessitate a significant amount of manual labor to calibrate and configure a vast number of design factors and tool settings. Unfortunately, examining just one design point can take a long time, as design procedures like PnR might take hours or even days for big circuits. To expedite hardware innovation, it is critical to reduce design costs by reducing the time required to acquire accurate QoR estimation and minimizing human supervision during the design tuning process.

Recent years have seen an increasing application of ML to accelerate the design process and reduce human engineering efforts and are believed to have great potential to address more critical challenges in both ASIC and FPGA designs. Specifically, there are three major directions in applying ML in design flow optimizations -- \textit{(1) Fast and accurate approximation via predictive modeling} -- ML can be used as a statistical technique that mines domain-knowledge from historical and existing data to forecast future or unseen outcomes w.r.t to specific algorithmic or mathematical objectives. With the recent progress in advanced ML algorithms and neural architectures, ML can be used to construct generic and accurate approximations for given objectives, which can significantly boost the design process \cite{ustun2019lamda,wu2022hybrid,DBLP:conf/dac/YuXM18,dai2018fast,ustun2020accurate,ren2023machine,pal2022machine,yin2023respect}. A well-calibrated ML predictive model can replace such heavy computations with a fast approximation.
\textit{(2) Flexible and versatile modeling} -- Unlike traditional statistical data analysis methods, modern ML techniques provide a wide range of modeling options to cover the complex FPGA design processes. On one hand, ML offers various predictive formulations that are essential to cover a large number of FPGA design challenges, e.g., classification, clustering, regression, generative modeling, etc (\cite{DBLP:conf/dac/YuXM18,dai2018fast,neto2021slap,neto2021read,yu2019painting,yu2018end}). On the other hand, modern variants of ML are able to handle versatile feature representations such as graphs, circuit imaging, functional behaviors, etc., and learn complex behaviors between those features and target metrics. \textit{(3) Minimizing human supervision} -- Leveraging ML in FPGA design minimizes human supervision in the design process in two directions. First, the traditional CAD tool R\&D process heavily relies on expert knowledge in FPGA design and CAD algorithms, and most heuristics are empirically developed with tremendous experimental efforts (\cite{yu2020flowtune,zhu2020exploring,peruvemba2021rl,hosny2019drills}). On the other hand, autonomous exploration and learning systems such as reinforcement learning mechanisms can significantly accelerate the exploration efforts with an intelligent self-guided agent.


{In this context, recent years have seen an increasing employment of machine learning (ML) techniques to enable autonomous design space exploration, {reducing manual efforts and boosting design closure \cite{kapre2015intime,ustun2019lamda,liu2013-ml-hls,pasandi2019approximate,ziegler2017ibm,li2016efficient,DBLP:conf/dac/YuXM18,hosny2019drills,ma2019high,yu2020decision} for ASICs \cite{ziegler2017ibm,li2016efficient,DBLP:conf/dac/YuXM18,hosny2019drills} and FPGAs \cite{kapre2015intime,ustun2019lamda,liu2013-ml-hls,schafer2019high,liu2019accelerating}, and acceleration formal verification solvers \cite{cherif2021kissat,wu2023gamora,yu2017fast}.} }These works have two main flavors: (i) focus on EDA tool parameters tuning, i.e., binary switches (e.g., remapping on/off), and multi-label switches; (ii) exploring the sequence of synthesis transformations in an iterative training-exploration fashion through Convolutional Neural Networks (CNNs) \cite{DBLP:conf/dac/YuXM18} and reinforcement learning \cite{hosny2019drills}. Also, the different proposed approaches rely in both offline \cite{ziegler2017ibm} and online datasets \cite{kapre2015intime,ustun2019lamda} have also been proposed. 
Still, the approaches proposed so far have the following limitations: 

\begin{itemize}

\item \textbf{Limited theoretical guarantees}. While ML-based systems have the potential to produce good results, there are no theoretical guarantees in terms of exploration bound and failure prediction.

\item \textbf{Lacking domain knowledge of synthesis algorithms}. Graph-based algorithms are the most widely used as fundamentals of logic optimization techniques. However, recent ML-based approaches consider synthesis algorithms or options as black-box implementations and have not conducted any graph algorithm characteristics in the learning approaches \cite{liu2013-ml-hls,ustun2019lamda,DBLP:conf/dac/YuXM18,hosny2019drills}.

\item \textbf{Limited transferability and flexibility}. It is known that ML models are highly limited to the problem space of the given dataset. Similarly in ML for synthesis, it is mostly limited to specific QoR objective(s) and challenging to transfer learned knowledge cross different designs \cite{ustun2019lamda,DBLP:conf/dac/YuXM18,hosny2019drills}.

\item \textbf{System integration overhead}. Due to the significant difference in back-end kernels and front-end user interfaces of EDA tools and ML frameworks, there is significant runtime and integration overhead while integrating existing ML techniques in EDA flows (e.g., TensorFlow used in \cite{DBLP:conf/dac/YuXM18,hosny2019drills,xie2018routenet,xu2019wellgan}).

\item \textbf{Challenges in end-to-end validations}. Existing works on logic optimization explorations are all evaluated at post-synthesis or post-mapping stages, without evaluating results at post physical design stage \cite{hosny2019drills,DBLP:conf/dac/YuXM18,zhu2020exploring,yu2020flowtune}, which can have significant impacts on the QoRs.

\end{itemize}

Therefore, to tackle the aforementioned flaws, in this work we present and thoroughly discuss previously presented machine learning techniques for logic synthesis, and propose an easily portable light-weight multi-armed bandit (MAB) approach for Boolean logic optimization, called \textit{FlowTune}. Past works have shown the effectiveness of the proposed framework to reduce the cardinality of Conjunction Normal Form (CNF) for Boolean Satisfiability (SAT), as well as its effectiveness to reduce the circuit directed-acyclic graph (DAG) size, before technology mapping \cite{yu2020flowtune}. In this work, we focus on technology dependent metrics, \textit{i.e.,} post-mapping. Thus, we present and discuss results for \textit{FlowTune} in different scenarios: STA-timing aware standard-cell (STD) technology mapping and post-FPGA placement and routing. 
Previous work have not focused on technology-dependent metrics, but this is fundamental to assess the effectiveness of ML-based approach for logic optimization. 
To make it possible, we integrate \textit{FlowTune} with two different logic optimization mechanisms, And-Inv-Graph (AIG) \cite{mishchenko:2006-dag,yu2016dag} and Majority-Inv-Graph (MIG) \cite{amaru2014majority}, and various design toolflows for end-to-end evaluation, such as ABC+LSOracle+VTR 8.0 \cite{neto2019lsoracle,luu2014vtr}, ABC+Yosys \cite{wolf2016yosys} with different back-ends such as Vivado, OpenFPGA \cite{tang2019openfpga}, and Cadence Genus, for STA analysis.
Specifically, the main contributions of this work include:

\begin{itemize}
    \item  A novel domain-specific bandit algorithm for sequential decision-making by leveraging domain knowledge of DAG-aware synthesis algorithms, with a detailed domain-knowledge illustration example. We show that handcrafted recipes for both AIG- and MIG-based synthesis lack a better area \textit{vs} performance trade-off. 
    
    \item The proposed framework has been implemented in ABC and LSOracle, and integrated with multiple open-source design toolflows for both ASICs and FPGAs evaluations, which enables end-to-end experimental evaluations to post-PnR stages.

    \item The proposed domain-specific bandit approach enables flexible and efficient exploration for logic optimizations with comprehensive experimental demonstrations, including Boolean minimization (depth and logic count), post-mapping optimization (delay and area), and post-PnR optimization (timing slacks, logic area, and routing area). We believe this is the first work that demonstrates the effectiveness of ML-guided synthesis exploration with complete PnR evaluations, using two different FPGA backend tools (VPR and OpenFPGA).
    
    \item We demonstrate the effectiveness of the domain-specific MAB approach in two different logic synthesis DAG representations, i.e., AIG-based and MIG-based logic synthesis, which is the first work that addresses MIG synthesis flow exploration in end-to-end settings.
    
    \item {FlowTune framework and its integration of multiple toolflows are released publicly \footnote{\url{https://github.com/Yu-Utah/FlowTune}}.}
\end{itemize}


\vspace{-3mm}
\section{Background}

This section presents useful notations for the reader, as well as reviews the search space while generating synthesis flows. We then present a motivational example on how delay and area may vary with synthesis flow, and how these gains are comparable w.r.t the state-of-the-art. 

\subsection{Notations and Search Space}\label{sec:space}

\noindent
\textbf{Definition 1 \textit{none}-repetition Synthesis Flow}: \textit{Given a set of unique synthesis transformations $\mathbb{S}$=\{$p_0$, $p_1$,..., $p_n$\}, a synthesis flow $\mathbb{F}$ is a permutation of $p_i$ $\in$ $\mathbb{S}$ performed iteratively.}

\textbf{Example 1:} Let $\mathbb{S}$=\{$p_0$, $p_1$, $p_2$\}. $p_i$ are the transformations in the synthesis tools and can be processed independently. Then, there are totally six flows available: 

\vspace{0mm}
\begin{equation*}
\small
\begin{aligned}
F_0 : p_0 \rightarrow p_1 \rightarrow p_2 ~~
F_1 : p_0 \rightarrow p_2 \rightarrow p_1 \\
F_2 : p_1 \rightarrow p_0 \rightarrow p_2 ~~
F_3 : p_1 \rightarrow p_2 \rightarrow p_0 \\
F_4 : p_2 \rightarrow p_0 \rightarrow p_1 ~~
F_5 : p_2 \rightarrow p_1 \rightarrow p_0
\end{aligned}
\end{equation*}

\textbf{Remark 1}: \textit{Let $f(n,1)$ be the upper bound number of possible flows, where $\mathbb{S}$ includes $n$ elements, and each transformation appears only once, such that:}

\vspace{-2mm}
\begin{align}
\small
f(n,1) = n!
\end{align}

The upper bound of $\mathbb{N}$ happens \textit{iff} all elements in $\mathbb{S}$ can be processed independently. In practice, there could be some constraints to be satisfied for processing these transformations. In this case, $\mathbb{N}$ will be smaller than $n!$. For example, given a constraint that $p_1$ has to be processed before $p_2$, the available flows include only $F_0$, $F_2$, and $F_3$.

\textbf{Definition 2 $m$-repetition Synthesis Flow (m$\geq$2)}: \textit{Given a set of unique synthesis transformations $\mathbb{S}$=\{$p_0$, $p_1$,...,$p_n$\}, a synthesis flow with $m$-repetition $\mathbb{F}_{m}$ is a permutation of $p_i$ $\in$ $\mathbb{S}_{m}$, where $\mathbb{S}_{m}$ contains $m$ $\mathbb{S}$ sets.} 

\textbf{Example 2:} Let $\mathbb{S}$=\{$p_0$, $p_1$\}. Each $p_i$ can be processed independently. For developing $2$-repetition synthesis flows, $\mathbb{S}_{2}$=\{$p_0$, $p_1$, $p_0$, $p_1$\}. The available flows include:

\vspace{-2mm}
\begin{equation*}
\small
\begin{aligned}
F_0 : p_0 \rightarrow p_0 \rightarrow p_1 \rightarrow p_1~~
F_1 : p_1 \rightarrow p_1 \rightarrow p_0 \rightarrow p_0 \\
F_2 : p_0 \rightarrow p_1 \rightarrow p_0 \rightarrow p_1~~
F_3 : p_1 \rightarrow p_0 \rightarrow p_1 \rightarrow p_0 \\
F_4 : p_0 \rightarrow p_1 \rightarrow p_1 \rightarrow p_0~~
F_5 : p_1 \rightarrow p_0 \rightarrow p_0 \rightarrow p_1 \\
\end{aligned}
\end{equation*}

\textbf{Remark 2}: \textit{Let $\mathbb{L}$ be the length of a synthesis flow. Given a  $m$-repetition $\mathbb{F}_{m}$ with $n$ transformations in $\mathbb{S}$, $\mathbb{L}$ = $n$$\times$$m$.}

The search space for $m$-repetition flows is a \textit{multiset} permutation problem. Hence, a closed formula can be derived to describe the search space of $m$-repetition flows. The search space of $m$-repetition flows with $n$ unique transformations is shown in Equation \ref{eq:space1}.

\textbf{Remark 3}: \textit{Let function $f(n,m)$ be upper bound number of available $m$-repetition flows with $n$ elements in $S$. $f(n,m)$ uniquely satisfies the following formula :}

\begin{align}
\small
\begin{aligned}
f(n,m) = \dfrac{(n\cdot m)!}{(m!)^n}
\end{aligned}
\label{eq:space1}
\end{align}

With the multiset permutation concept, Yu et al. \cite{DBLP:conf/dac/YuXM18} generalized the formula to describe the search space for any type of \textit{m-}repetition flows. {Let $n$ be the number of unique transformation, the $M$-repetition flows, $M$=\{$m_{0}, m_{1}, ...,m_{n-1}$\}, where $m_i$ is the number of repetitions of the $i^{th}$ transformation}. The total number of possible flows is shown in Equation 2. 

\textbf{Remark 4}: {\textit{Let function $f(n,\mathbb{L},\{m_{0}, m_{1}, ...,m_{n-1}\})$ be upper bound number of flows with $n$ elements in $S$}. The $i^{th}$ element in $S$ appears $m_i$ times. The function uniquely satisfies the following formula :}

\vspace{-2mm}
\begin{align}
\scriptsize
f(n,\mathbb{L},\{m_{0}, m_{1}, ...,m_{n-1}\}) = \dfrac{(m_0+m_1+\cdot\cdot\cdot m_{n-1})!}{(m_0!)(m_1!)\cdot\cdot\cdot(m_{n-1}!)}
\end{align}

\textbf{Remark 5}: {\textit{Let {$\mathbb{L}$ be the length of the type of flows with upper bound $f(n,\mathbb{L},\{m_{0}, m_{1}, ...,m_{n-1}\})$, $\mathbb{L}$=$\sum_{i=0}^{n-1}m_i$.}}}

Whereas the work in \cite{DBLP:conf/dac/YuXM18} constraints the search-space to \textit{m-}repetitions flow, with the upper bound given by Equation 2, in this work we approach this problem in a more general way, where each repetition might have a different number of repetitions. In this case, the theoretical upper bound is given by Equation 3.

\subsection{Motivating Example}\label{sec:motivating-example}

\begin{figure}[!h]
\centering
\includegraphics[width=0.44\textwidth]{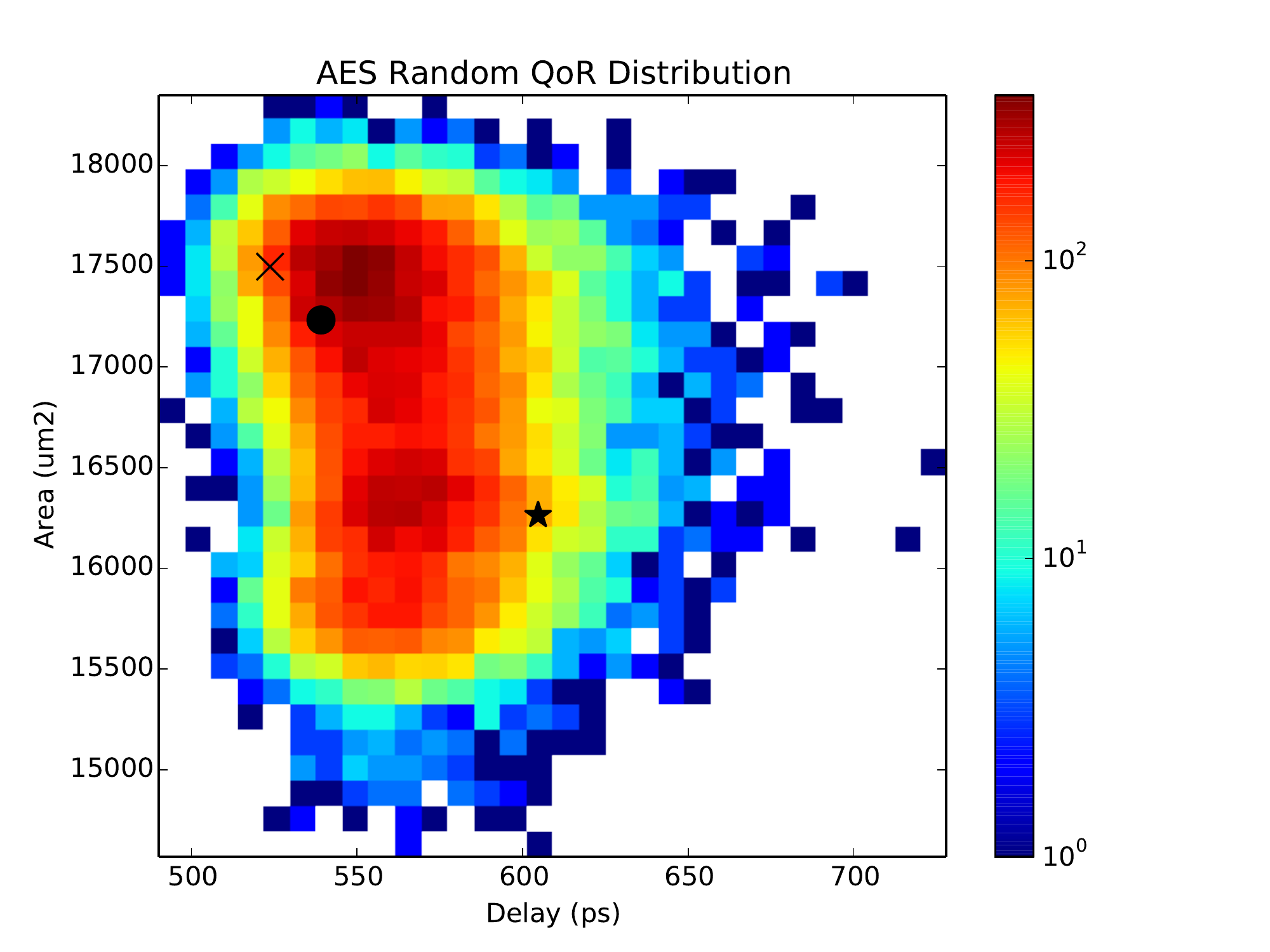}
\caption{Evaluation of three default flows, \textit{resyn ('o'), resyn2 ('x')} and \textit{resyn3 (the star)} using 128-bit AES design. The heatmap includes the QoR of the 50,000 random flows.}
\label{fig:default_flow_compare}
\end{figure}

\begin{figure}[!h]
\centering
\includegraphics[width=0.44\textwidth]{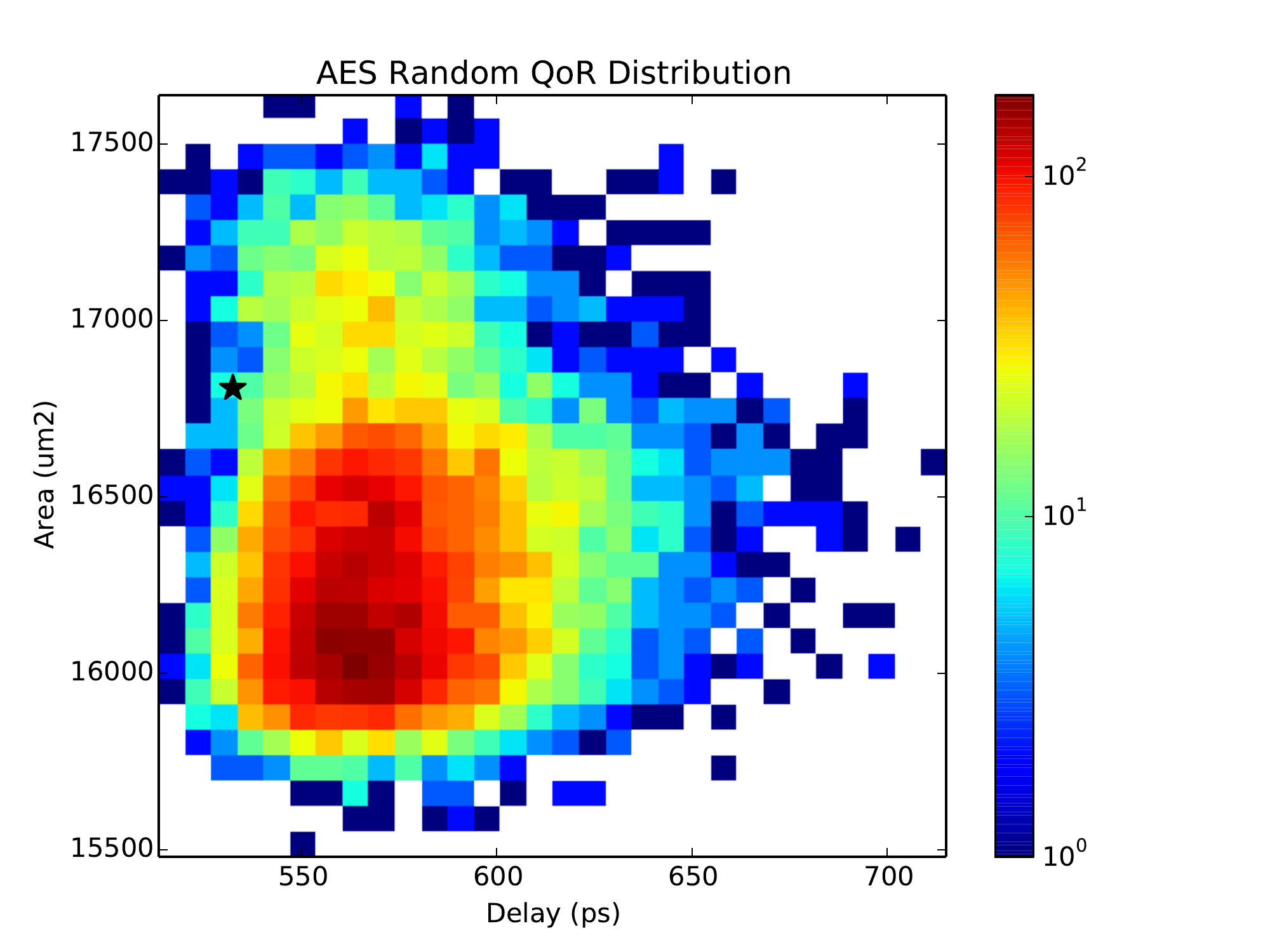}
\caption{Evaluation of the LSOracle default flow (black star) using 128-bit AES design. The heatmap includes the QoR of the 50,000 random flows.}
\label{fig:lso_flow_compare}
\end{figure}

We provide two motivating examples using open-source logic synthesis framework ABC \cite{mishchenko2010abc} and LSOracle \cite{neto2019lsoracle}, in which the backend data structures are AIGs and MIGs, respectively. 
The experimental setups for the motivational examples are as follows:

\noindent
\textbf{ABC \cite{mishchenko2010abc}} -- We use the following set of commands: $\mathbb{S}$=\{balance, restructure, rewrite, refactor, rewrite -z, refactor -z\} ($n$=6); the elements in  $\mathbb{S}$ are logic transformations in ABC\footnote{\small The names of these transformations are the same as the commands in ABC.} that can be processed independently. We take as input a 128-bit Advanced Encryption Standard (AES) core\cite{opencore-web} and generate 50,000 unique $4$-repetition flows are generated by random permutations of  $\mathbb{S}_{4}$ ($m$=4, $n$=6, $\mathbb{L}$=24). The delay and area of these flows are obtained after technology mapping using the ASAP 7nm predictive standard-cell library.  

\noindent
\textbf{LSOracle\cite{neto2019lsoracle}} -- We use the following transformations: $\mathbb{S}$=\{rewrite, resub, refactor\} ($n$=3). That is because these are the main transformations used in the highest MIG-based transformation in LSOracle. We take as input a 128-bit Advanced Encryption Standard (AES) core\cite{opencore-web} and generate 50,000 unique $4$-repetition flows are generated by random permutations of  $\mathbb{S}_{4}$ ($m$=3, $n$=9, $\mathbb{L}$=27). We chose to have a flow of length 27 as it is a multiplier of the default LSOracle flow, which has 9 transformations by default. To compare, we run the default LSOracle flow 3$\times$. The delay and area of these flows are obtained after technology mapping using the ASAP 7nm predictive standard-cell library.

The QoR distributions for the AES can be seen in Figures \ref{fig:default_flow_compare},\ref{fig:lso_flow_compare}, and several important observations can be drawn from it, which highlight the main motivations of this work:
    \vspace{0mm}

\begin{itemize}
    \item (1) Given the same set of synthesis transformations, the QoR is very different using different flows, for both AIG and MIG. For example, for the AIG case, the delay and area of AES design produced by the 50,000 flows have up to 40\% and 90\% difference, respectively. As for the MIG-based flow, the variance of delay and area are up to {39.27\% and 13.97\%}, respectively. 
    \vspace{0mm}
    \item (2) The search space of the synthesis flows is large. According to Remark 3, the total number of available $4$-repetition flows with $n=6$ independent synthesis transformations is more than $10^{16}$ (considering our AIG example). Discovering the high-quality synthesis flows with human-testing among the entire search space is unlikely to be achieved. Same holds for MIGs.
        \vspace{0mm}
        \vspace{0mm}
\end{itemize}


As it is possible to observe, we position the three most widely used default flows provided in ABC, including \textit{resyn, resyn2} and \textit{resyn3}, with respect to the QoR achieved with the random flows. Whereas \textit{resyn} provides the best QoR among these three default flows, its result performs $>$10\% worse than the best flows of the 50,000 random flows, in both delay and area. Specifically, there are more than 25,000 random flows that perform better than all the default flows. This means that for every two random permuted flows, one of them is likely to over-perform the expert-developed flows for this design. 
Similarly, we show that a high-effort and well expert-tuned flow in LSOracle flow MIGs still have much room for improvement. In particular, the default flow presents a good delay overall, but with room for achieving a better area ($\approx$7\% worse area than random flows with similar delay). 
These motivational examples show that automatic flow generation is an important direction for both AIG- and MIG-based synthesis flow generation. This, along with the large search space, provides the main motivation for this work. It is also important to note that a given flow performs differently on different designs. For example, high-quality flows for our AES design could perform poorly on other designs \cite{DBLP:conf/dac/YuXM18}.

\section{Approach}\label{sec:approach}

\subsection{On the Impact of DAG-Aware Synthesis Algorithms}\label{sec:algorithm}

\begin{figure}[t]
\begin{minipage}{.24\textwidth}
\scriptsize
\begin{algorithmic}[1]
 \Procedure{dag-aware synthesis}\;\\
  $G(\mathcal{V},\mathcal{E})$ $\gets$ circuit\;\\
 \For{$v \in \mathcal{V}$}{
 \If{\color{red} transformable($v$)}{
  {\color{red} apply transformation to $v$}\; \\
  update $G(\mathcal{V},\mathcal{E})$\;
  }
 }
 \EndProcedure
\end{algorithmic}
\end{minipage}%
\begin{minipage}{.25\textwidth}
 \centering
\includegraphics[width=1\textwidth]{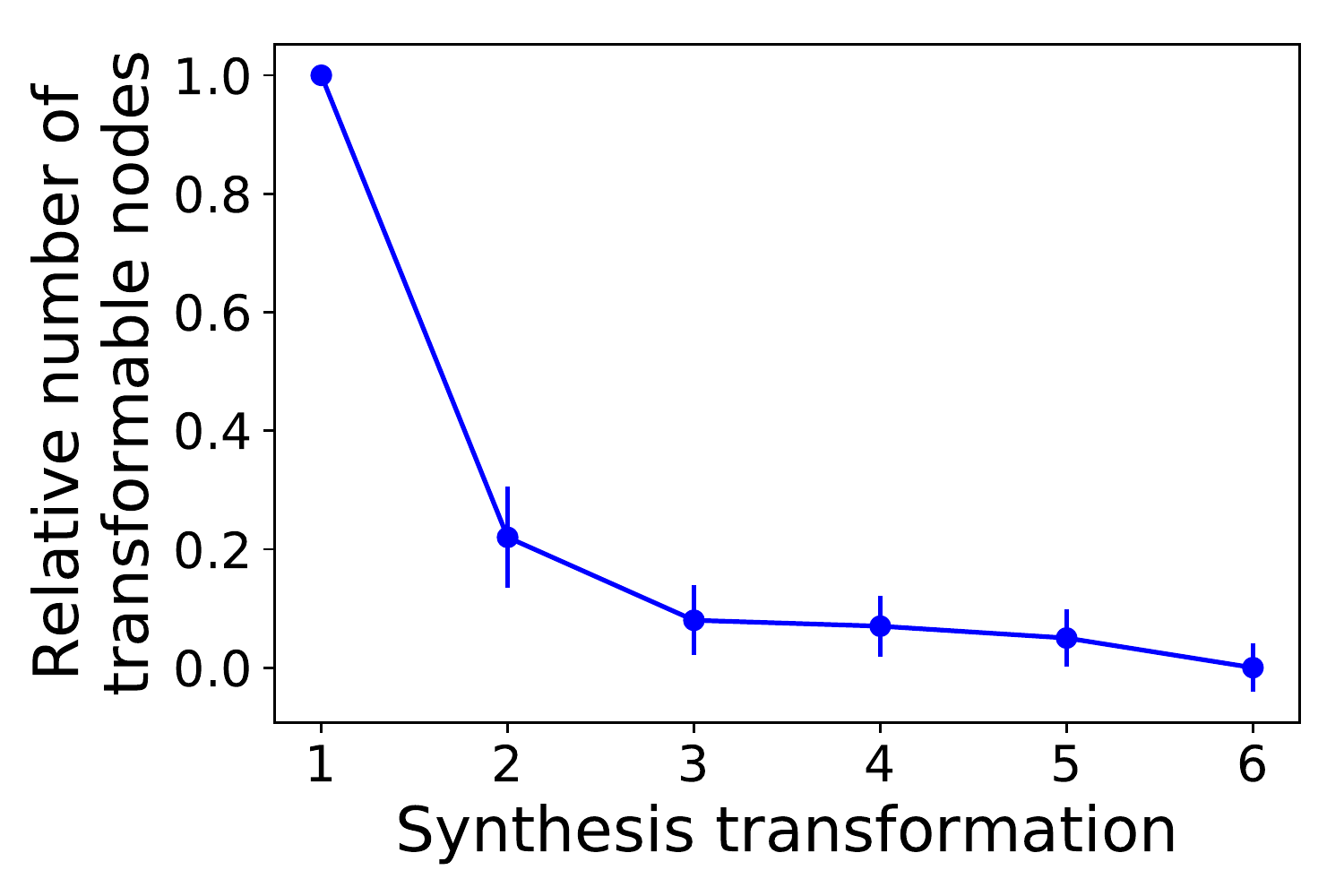}
\end{minipage}
\caption{Illustration of DAG-aware synthesis algorithm. And, the {relative number} of AIG nodes that are effectively executed in each transformation of the synthesis flows, using 100 random flows with six transformations used in \cite{DBLP:conf/dac/YuXM18,hosny2019drills}.}
\label{fig:motivation}
\end{figure}

The most efficient algorithms that optimize the Boolean networks are directed acyclic graph (DAG) aware based Boolean synthesis algorithms \cite{mishchenko:2006-dag,wolf2016yosys}, which are widely used in both open-source tools \cite{mishchenko2010abc,wolf2016yosys,luu2014vtr, neto2019lsoracle} and industrial tools \cite{cunxiyu:dac16,stok1996booledozer,DBLP:conf/iccad/YuCSC17,amaru2014majority}. Specifically, this work focuses on optimizing the synthesis flows that comprise DAG-aware synthesis algorithms and heuristics, targeting AIG- and MIG-based flows. Thus, to understand how effective each synthesis transformation (algorithm) is in the synthesis flows, we analyze the basic graph operations in DAG-aware algorithms. We showcase the number of transformed nodes for an AIG-based flow, running 6 different transformations  {(bracket 4 in the pseudo-code of the Algorithm presented in Fig. \ref{fig:motivation})}. With that, we can estimate the effectiveness of the algorithm for a given DAG (a circuit). Hence, to understand how effective each synthesis transformation (algorithm) is in the synthesis flows, we monitor the number of transformed nodes of all transformations using 100 random flows. The selected ABC synthesis transformations are the same six transformations as in \cite{DBLP:conf/dac/YuXM18,hosny2019drills}, namely $\mathbb{S}$=\{balance, restructure, rewrite, refactor, rewrite -z, refactor -z\}. We collect these results for six VTR benchmarks, as shown in Table \ref{tbl:benchmark}, and plot the average, max, min number of relative transformable nodes. While we plot the average over six designs, we observed a similar trend in all the designs. 

The analysis results are shown in Figure \ref{fig:motivation}, where the y-axis represents the relative number of transformed nodes of each transformation, and the x-axis shows the steps of the synthesis flows. For example, given a random \textit{none}-repetition flow composed by the transformations available in $\mathbb{S}$, assume the first transformation is applied to $\sim$1,000 nodes in the {original} graph and the second transformation is applied to $\sim$200 nodes in the {updated} graph. In which case, the relative percentage of the first two transformations of this example is denoted as $1$ and $0.2$. Therefore, in Figure \ref{fig:motivation}, the relative number of transformed nodes of all randoms flows start with $1$. As there are 100 random permuted flows for six designs in this analysis, the error-bars are used to indicate the upper/lower bound of transformable nodes among all the random generated flows. MIG-based transformations have a similar trend, with a particularity: the number of relative transformable nodes for the first two transformations tend to be closer. {Our intuition is that the input for MIG-based synthesis is an AIG data structure, where the nodes have at least one input as constant (true or false)}. Thus, it limits the number of applicable MIG-based transformations on the first iteration. On the other hand, the second iteration has more majority nodes, so there are still a good amount of transformable nodes for MIG-based algorithms. Thus, the observations done for AIGs also hold for MIGs.

There are two important observations can be concluded empirically from the experiments: 

\begin{itemize}

\item Observation (1) \textit{The earlier transformations selected in a given synthesis flow have the highest impacts to the Boolean optimization problem w.r.t DAG-aware logic transformations.}

\end{itemize}

In Figure \ref{fig:motivation}, we can see that for any random flow, the third to sixth transformations in the flow are applied to less than 10\% nodes compared to the first transformation. In this particular example, we observe many cases that the fifth and sixth transformations do not detect any transformable node, regardless of the permutations. 

\begin{itemize}

\item Observation (2) \textit{Earlier transformations in the synthesis flow dominates the performance of Boolean optimizations since the transformable nodes for the rest of the flow are determined by earlier transformations (local minima).}

\end{itemize}

That being said, starting a synthesis flow with an unsuccessful transformation will almost certainly result in an inefficient flow, regardless of how the other optimizations are permuted. This is also true for MIGs, because the first transformation sets the topology of the real MIG graph, influencing the efficacy of subsequent MIG-based transformations. As a result, having the first transformation bias the flow is equivalent to solving a non-convex optimization problem with an initialization that always converges at a suboptimal local optimum. We may conclude from the suggested experiment that the performance of synthesis flows for DAG-based transformations is dominated by the first transformation in the flow. With these two observations in mind, we propose a domain-knowledge MAB approach for automatic logic synthesis flow generation. 


\noindent
\textbf{Example 1:} An illustrative example that demonstrates the importance of the extracted algorithmic domain knowledge presented in Figure \ref{fig:motivation} is shown in Table \ref{tbl:ex1}. Two DAG-aware synthesis transformations from ABC are selected to build two different flows $F_0$ and $F_1$ and are applied to design \texttt{bfly} from VTR \cite{luu2014vtr} benchmark. We focus on comparing the transformed AIG nodes (\#TNodes) of \texttt{rewrite} and the final number of AIG nodes. This is because \texttt{rewrite} performs a technology-independent rewriting algorithm of AIG, which offers the main reductions of \#AIG in this example. We can see that \textbf{(1)} \textit{The transformations at the early stage of the flows have more impacts than the transformations at the late stage.} For example, in both flows, first \texttt{rewrite} successfully applies to more than 7$\times$ \#AIG than the second \texttt{rewrite}. \textbf{(2)} \textit{The choice of early transformations has significant impacts on the performance of the flow}. Early transformations have more impacts on the Boolean network, so the DAG structure could change dramatically at the early stage compared to the late stage. For example, if we apply \texttt{balance} first, the first \texttt{rewrite} in $F_0$ applies to 951 nodes. On the other hand, without \texttt{balance}, the first  \texttt{rewrite} can be applied to 1764 nodes at ($F_1$). Similar results can be observed from the remaining two \texttt{rewrite}.

\begin{table}[t]
\caption{Example of the algorithmic domain knowledge presented in Figure \ref{fig:motivation} using ABC synthesis transformation \textit{rewrite} rw and \textit{balance} (b) with design {bfly} from VTR 8.0 \cite{luu2014vtr} benchmark. Note that rw is technology-independent rewriting of the AIG which offers the main AIG reductions in $F_0$ and $F_1$. \#TNodes = Number of AIG nodes transformed by the corresponding AIG transformation.}
\begin{tabular}{cccccccc}
\textit{$F_0$} & \textit{b} & \textit{rw} & \textit{b} & \textit{rw} & \textit{b} & \textit{rw} & Final \#AIG \\
\textit{\#TNodes} & 817 & \textbf{951} & 825 & \textbf{169} & 831 & \textbf{64} & 26339 \\ \hline\hline
\textit{$F_1$} & \textit{rw} & \textit{b} & \textit{rw} & \textit{b} & \textit{rw} & \textit{b} & Final \#AIG \\
\textit{\#TNodes} & \textbf{1764} & 824 & \textbf{290} & 834 & \textbf{90} & 832 & \textbf{26182}
\end{tabular}
\label{tbl:ex1}
\end{table}

\noindent
\textbf{Example 2:} Now, we show the importance of getting the right transformation for MIG-based synthesis, according to our discussion on MIG-based synthesis trends in Figure \ref{fig:motivation}. The experiment is depicted in Table \ref{tbl:ex2}. 
We select two area-oriented (nodes reduction) transformation to this end, so we aim to highlight the effectiveness of picking a right transformation at the 1st iteration to the overall circuit area (DAG size) reduction. 
Similarly to the previous example, we use two transformations (resub and refactor) to build two different flows $F_0$ and $F_1$. These flows are applied to the same design (\texttt{bfly}) as the previous example. Note that we do not intend to compare the effectiveness of AIG vs MIG based synthesis with this experiment, but rather show the importance of leveraging domain-knowledge for automatic flow generation. 
As in the previous example, we can see that \textbf{(1)} \textit{The transformations at the early stage of the flows have more impacts than the transformations at the late stage.} Also, as we discussed, we can see that for MIG-based, effective flows tend to have a similar number of transformable nodes for the first two transformations, as the first transformation actually defines the actual MIG structure.  
\textbf{(2)} \textit{The choice of early transformations has significant impacts on the performance of the flow}. In this case, the same observation as we did for AIGs hold for MIGs, and the early transformations have more impacts. 

\begin{table}[t]
\caption{Example of the algorithmic domain knowledge presented in Figure \ref{fig:motivation} using MIG-based synthesis transformation \textit{refactor} (rf) and \textit{resub} (rs) with design {bfly} from VTR 8.0 \cite{luu2014vtr} benchmark. We use these transformations as they are area-oriented within the LSOracle flow, and can highlight the importance of picking the right transformation w.r.t the effectiveness of transformable and reduced nodes. \#TNodes = Number of MIG nodes transformed by the corresponding MIG transformation.}
\begin{tabular}{cccccccc}
\textit{$F_0$} & \textit{rf} & \textit{rs} & \textit{rf} & \textit{rs} & \textit{rf} & \textit{rs} & Final \#MIG \\
\textit{\#TNodes} & 984 & 931 & 87 & 44 & 3 & 1 & \textbf{26659} \\ \hline\hline
\textit{$F_1$} & \textit{rs} & \textit{rf} & \textit{rs} & \textit{rf} & \textit{rs} & \textit{rf} & Final \#MIG \\
\textit{\#TNodes} & 1402 & 377 & 74 & 15 & 4 & 1 & 26779
\end{tabular}
\label{tbl:ex2}
\end{table}

\vspace{-3mm}
\subsection{Domain-specific MAB Formulation}\label{sec:mab}

An online program must choose from a set of strategies in a sequence of $n$ trials to maximize the overall payout of the chosen strategies in a multi-armed bandit problem. Such challenges require that a set quantity of resources must be distributed among accessible options in such a way that the expected gain of a given objective is maximized. These are the most important theoretical tools for modeling the exploration-exploitation trade-offs that are inherent in sequential decision-making under uncertainty. Specifically, MAB process is represented as a tuple of $<\mathcal{A},\mathcal{R}>$, where:

{\it
\begin{itemize}
\item $\mathcal{A}$ is a known set of available choices (arms).
\item At each time step $t$, an action $a_t$ is triggered by choosing with one of the choice $a_t \in \mathcal{A}$. 
\item $\mathcal{R}$ is a reward function and $\mathcal{R}^{a_t}$ is the reward at time step $t$ with action $a_t$.
\item The objective of bandit algorithm is to maximize $\textstyle \sum^{t}_{i} \mathcal{R}^{a_t}$
\end{itemize}
}

In a classic MAB sequential decision-making environment, the available decisions at time step $t$ are considered as arms. For example, considering the synthesis flow exploration problem, the selected synthesis transformations will be the set of arms $\mathcal{A}$. Let $\mathcal{A}$ include eight unique transformations $\mathcal{A}$=\{\texttt{resub,rewrite,...,refactor}\}. Let $\mathcal{R}$ be the number of AIG nodes that have been reduced by applying the transformations, such that the objective of the bandit algorithm is to maximize $\textstyle \sum^{t}_{i} \mathcal{R}^{a_t}$, i.e., minimize the number of AIG nodes. Let $F$ be a decision sequence, where $F$=\{$a_0, a_1,..., a_n$\}, $a_t \in \mathcal{A}$. This decision sequence $F$ is a synthesis flow. %
A brute-force approach to identifying the optimum flow that maximizes the benefit $F$ performs enough rounds with each transformation (each arm) to determine the true probability of reward. Unfortunately, a brute-force method like this is impractical. In this situation, the bandit algorithm's principal goal is to gather enough data to make the best overall decisions. According to prior plays, the best-known option is chosen during exploitation. The exploration phase will look into unknown choices inside the search space in order to gain more information and close the gap between estimated and true reward function probabilities. This process is analogous to reinforcement learning for synthesis flow exploration without internal states, such as a snapshot of current Boolean network statistics. Despite the fact that MAB sequential decision making is extensively used, it needs to be rethought to fit in with logic synthesis flow exploration. In particular, according to the two observations in Section \ref{sec:algorithm}: 

\begin{itemize}

\item Since the first synthesis transformation dominates the flow, the first action in exploration will dominate the true reward distribution $\mathcal{R}^{*}$ and the exploitation reward distribution $\mathcal{R}$. In other words, the initialization of the bandit algorithm dominates the gap between $\mathcal{R}^{*}$ and $\mathcal{R}$.  

\item While considering each transformation as an arm, each action corresponds to applying one synthesis transformation to the logic graph. Unlike the classic MAB problem that $\mathcal{R}^{*}$ is fixed over time, $\mathcal{R}^{*}$ in synthesis flow exploration changes at each time step, as the logic graph is updated by the transformation.

\end{itemize}

Therefore, to gather the domain knowledge of synthesis algorithms discussed in Section \ref{sec:algorithm}, we propose a novel MAB environment by re-defining the arms and actions, which fits the logic synthesis flow exploration problem. Thus, let $P(\mathcal{X})$ be a random permutation function over a set of decisions $\mathcal{X}$. Let $P(x|\mathcal{X})$ be a random permutation function over the set $\mathcal{X}$, $x \in \mathcal{X}$, such that $P(x_i|\mathcal{X})$ is a random permutation with $x_i$ always being the first element in the permutation, $P(x_i|\mathcal{X}) \in P(\mathcal{X})$. We define $P(x|\mathcal{X})$ to be the arms in the MAB environment, such that $\mathcal{A}$=\{$P(x_0|\mathcal{X})$,~$P(x_1|\mathcal{X})$,~...,~$P(x_n|\mathcal{X})$~\}, where $n$ is the number of available decisions in the exploration problem. Specifically, $n$ corresponds to the number of available synthesis transformations. \textbf{Unlike using traditional MAB algorithms, an action $a_t$ at time $t$ is a sampled permutation from $P(x_i|\mathcal{X})$.} In other words, $a_t$ is a \textit{multiset} over the set $\mathcal{X}$. Let $Q(a_t)$ be the action value that is obtained by applying $a_t$ to a given Boolean network at $t$ time step, the reward is $r_t$
\[
r_t = Q(a_t) - Q(a_{t-1}) \implies Q(P(x_i|\mathcal{X})) - Q(P(x_j|\mathcal{X}))
\]
where the $i^{th}$ arm is played at $t$ time step and $j^{th}$ arm is played at $t-1$ time step, {and $Q(a) = \mathop{\mathbb{E}}(p|a)$ calculates the mean reward (e.g., number of AIGs minimized) from the actions at $t$ step over the estimated winning probabilities $p$}. Finally, we use upper confidence bound (UCB) bandit algorithm as the agent, such that $a_t$ is chosen with estimated upper bound $U_t(a)${, where setting the probability of the true mean being greater than the UCB to be less than or equal to $t^{-1}$, also known as UCB1 algorithm \cite{auer2002finite}. The adopted upper bound can be calculated using \textit{Lai and Robbins} theorem \cite{lai1985asymptotically} and the Hoeffding’s inequality. According to the Hoeffding’s inequality the probability of choosing an action exceeds upper confidence bound (UCB) is bounded by
\begin{equation}
    P[Q(a) > Q_{t}(a) + U_{t}(a)] \leq e^{-2N_{t}(a)U_{t}(a)^{2}}
\end{equation}

In other words, using Hoeffding’s inequality to assign an upper bound to an arm’s mean reward where there is high probability that the true mean will be below the UCB assigned by the algorithm. 

Therefore, while solving $U_t(a)$ in UCB,
\begin{equation}
    e^{-2N_{t}(a)U_{t}(a)^{2}} = p \Rightarrow U_{t}(a) = \sqrt{\frac{-log~p}{2N_{t}(a)}}
\end{equation}
In this case, we can see that reducing the probability of $p$ will increase the rewards from UCB. Setting the probability $p$ of the true mean being greater than the UCB to be less than or equal to $t^{-4}$, a small probability that quickly converges to zero as the number of rounds $t$ grows, which is also known as UCB1 algorithm. However, we find that setting $p=t^{-1}$ is sufficient based on our empirical studies, which results in 
\cite{auer2002finite}.

\begin{equation}
    \sqrt{\frac{-log~p}{2N_{t}(a)}} = \sqrt{\frac{log~t}{2N_{t}(a)}} = \sqrt{\frac{t}{2N_{t}(a)}}, ~~ \text{let} ~ p=t^{-1}
\end{equation}
Thus, we get the following:

\begin{equation}
    a_{t} = \underset{a \in \mathcal{A}}{\mathrm{argmax}}\, \left[ Q(a) + U_{t}(a) \right], ~~ U_{t}(a) = \sqrt{\frac{log~ t}{2N_{t}(a)}}
\end{equation}


}


The performance of a multi-armed bandit algorithm is often evaluated in terms of its regret, defined as the gap between the expected payoff of the algorithm and that of an optimal strategy. In this work, the number of regrets equals to the number of synthesis flows that have been evaluated in the synthesis tool. Using the UCB algorithm, an asymptotic logarithmic total regret $L_t$ is achieved \cite{auer2002finite}:

\[
 \lim_{t \to \infty} L_t = 2log~t~\sum \Delta_{a}
\]
where $\Delta_{a}$ is the differences between arms in $\mathcal{A}$.

\begin{figure}[b]
\centering
\includegraphics[width=0.4\textwidth]{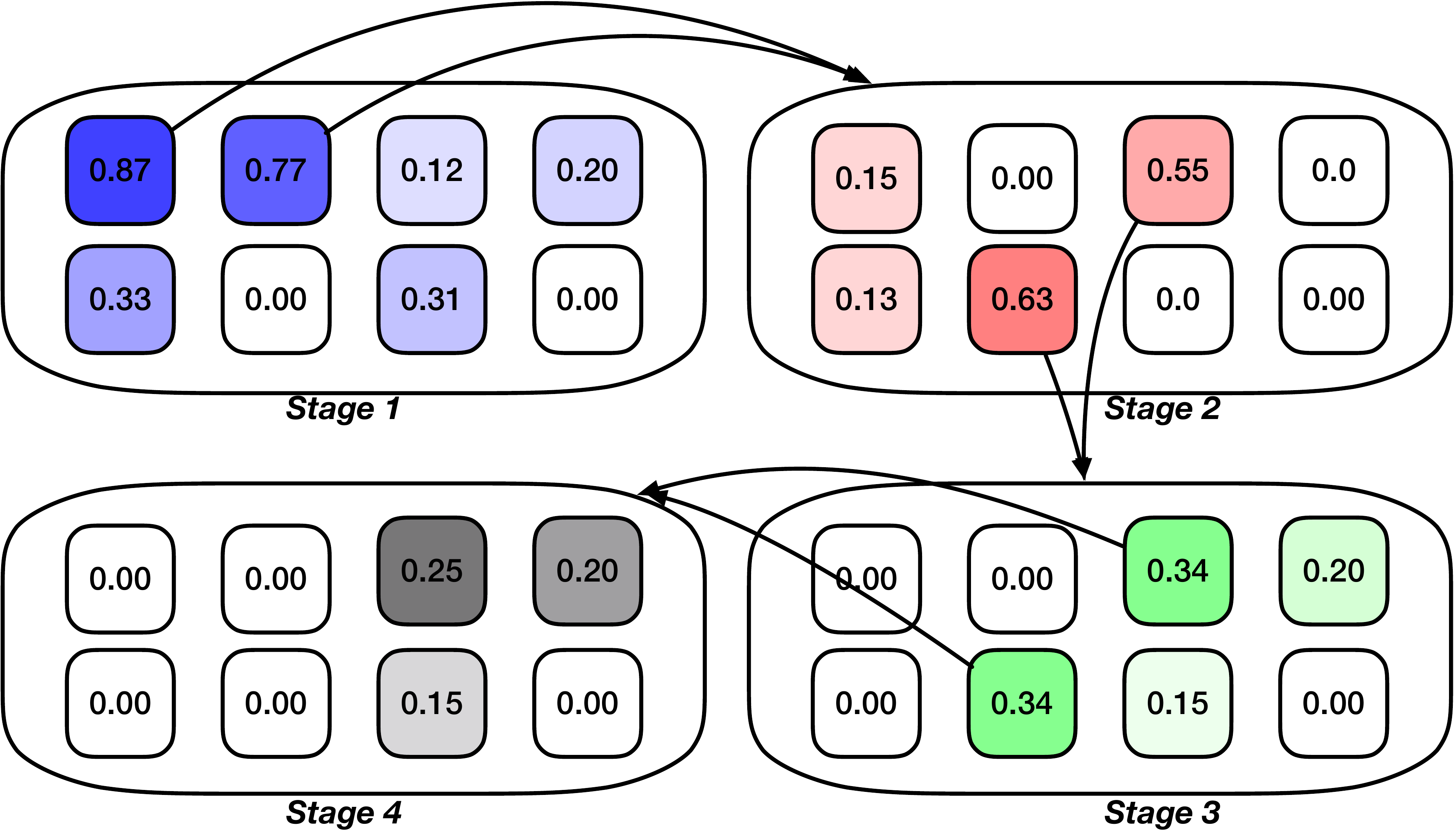}
\caption{Illustration of the proposed multi-stage bandit approach with four stages.}
\label{fig:multistage}
\vspace{-2mm}
\end{figure}

\subsection{Improving Convergence with Multi-stage Bandit}

While the preceding section's technique relies on optimistic initialization, we suggest a multi-stage bandit to further improve convergence. We can see that the single-stage approach may be used to longer synthesis flows, with each transformation being performed numerous times, based on the formulation in the previous section. However, increasing the length of the sequences leads to a significant increase in the number of explorations required to close the gap between the optimistic reward distribution $\mathcal{R}^*$ and exploitation reward distribution $\mathcal{R}$. Moreover, the optimistic reward distribution $\mathcal{R}^*$ changes as the synthesis transformations are applied to the logic circuit, since the graph structure is modified iteratively. Finally, while synthesis transformations are less successful at late time steps, fine-grain exploration can have a significant impact on EDA flows later on, such as technology mapping and gate sizing.

\begin{figure*}[t]
\centering
\includegraphics[width=0.85\textwidth]{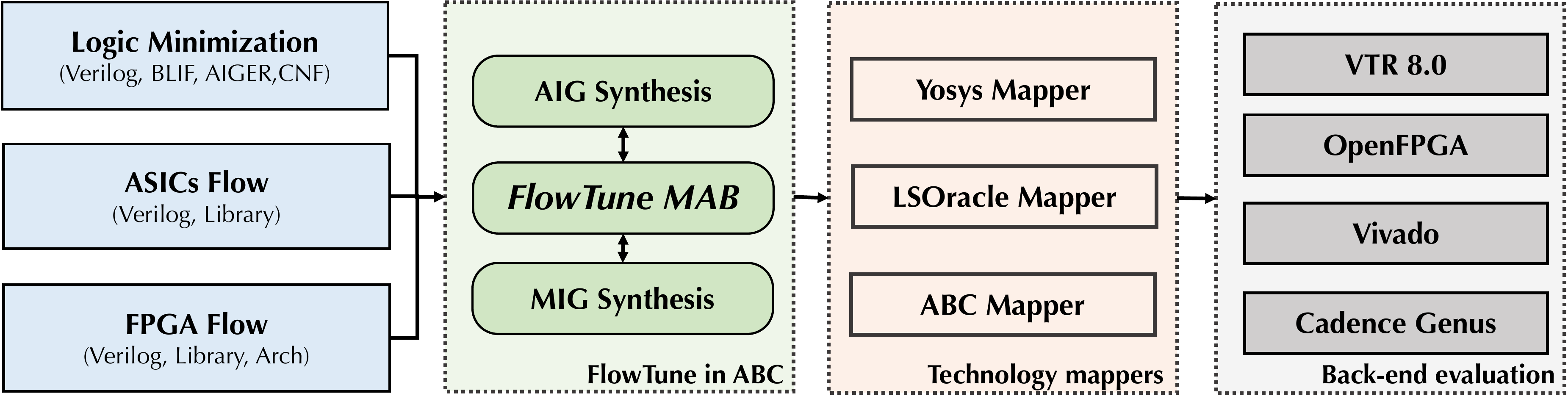}
\caption{Overview of the proposed end-to-end MAB synthesis system -- Using ABC front-end, our system accepts technology mapped netlist, Boolean logic netlist, and LUT-netlist. The system is also integrated with VTR 8.0 and Yosys, which enable synthesis optimization for a large range of objectives for designing ASICs and FPGAs, and formal verification tools.}
\label{fig:system_overview}
\end{figure*}

In this context, we introduce the proposed MAB algorithm using a four-stage example shown in Figure \ref{fig:multistage}. Each stage in the multi-stage approach uses the same domain-specific bandit algorithm described in Section \ref{sec:algorithm}. Within each stage, the MAB algorithm is restricted to a fixed number of iterations $m$. Once the first stage is completed, the exploitation reward distribution $\mathcal{R}^1$ is updated (stage 1 in Figure \ref{fig:multistage}), in which a higher rate indicates a higher chance of gaining reward by playing that arm. After $m$ iterations in the first stage, the best-explored synthesis flow will be applied to the input logic circuit, and the synthesized circuit will be the input circuit for the next stage. Rather than starting a new MAB agent with a uniform distribution, we start the second stage using the reward distribution of the first stage's top two arms. For example, in Figure \ref{fig:multistage}, $\mathcal{R}^{1}_{a_0}$ and $\mathcal{R}^{1}_{a_1}$ will be merged and used as the initial reward distribution for the second stage, where $\mathcal{R}^{1}_{a_0}, \mathcal{R}^{1}_{a_1} \in \mathcal{R}^{1}$. This procedure will continue until the $s$ stages have been completed. As we can see, the total number of explorations is $s \cdot m$. In this work, we have explored five different options for ($s,m$), while maintaining the total number of iterations identical.

\noindent
\textbf{Example 2}: We present an illustrative example of the aforementioned domain-specific MAB and the multi-stage bandit algorithm for exploring synthesis flows using the following ABC transformations: \texttt{rw, b, rf,} and \texttt{resub}. Let $\mathcal{X}$ be \{\texttt{4$\times$rw, 4$\times$b, 4$\times$rf, 4$\times$resub}\}. Let the number of stages for exploration be four, such that $\mathcal{X}_{0,1,2,3}=$\{\texttt{rw, b, rf, resub}\}. At first MAB stage, arms $\mathcal{A}_0$
\vspace{-1mm}
\begin{align}
\tiny 
\mathcal{A}_0=\{P(rw|\mathcal{X}_0), 
P(b|\mathcal{X}_0),
P(rf|\mathcal{X}_0),
P(resub|\mathcal{X}_0)~\}
\end{align}
are explored with the proposed MAB algorithm presented in Section \ref{sec:algorithm}. The reward $r_t$ is calculated based on the target objective, e.g., number of reduced AIG nodes compared to the existing best actions. Next, Stage 1 terminates after a fixed number of iterations, and returns the arm(s) with the highest reward value. For the sake of simplicity, this example only includes the arm with the highest reward for stage 2. Let us assume $P(rw|\mathcal{X}_0)$ and $P(rf|\mathcal{X}_0)$ return the highest reward at stage 1. We then define $\mathcal{A}_0^*=\{P(rw|\mathcal{X}_0),P(rf|\mathcal{X}_0)\}$, the arms for second stage, $\mathcal{A}_1$, are updated as follows:
\begin{align}
\tiny 
\mathcal{A}_1=\{\mathcal{A}_0^* \frown  P(rw|\mathcal{X}_1), ~...,\mathcal{A}_0^* \frown  P(resub|\mathcal{X}_1)~\}
\end{align}
where the sample from each arm in $\mathcal{A}_1$ will be a concatenation of two actions from $\mathcal{A}_0^*$ and $P$. For stage 3, we simply replace $\mathcal{A}_0^*$ with $\mathcal{A}_1^*$, which will be the highest reward arms from $\mathcal{A}_1$. Finally, note that the subsets $\mathcal{X}_{0,1,2,3}$ are not necessary to be equally defined. For example, we can define $\mathcal{X}_0$= \{2$\times$\texttt{rw,b,rf,resub}\}, $\mathcal{X}_{1,2}=$\{\texttt{rw, b, rf, resub}\}, and $\mathcal{X}_3$=\{\texttt{b,rf,resub}\}.

\section{FlowTune Framework}

\subsection{Initialization}

As we discussed, the initialization step is crucial for MAB-based exploration performance. We leverage the domain knowledge of DAG-aware synthesis algorithms in our initialization stage. Specifically, for our multi-stage MAB exploration approach, we initialize the reward value for each arm in the first stage using the total number of transformable nodes by sampling each arm. While as demonstrated in Example 1, the total number of transformable nodes for a sequence of transformations highly depends on the first transformation, the initialization involves only one sampling of each arm. More importantly, this scheme significantly reduces the runtime of initialization for objectives such as technology mapping, since counting the number of transformable nodes does not require the actual mapping process. The initialization process is used at the beginning of each stage. To further improve the speed of initialization, we implement a parallel sampling function using OpenMP library \cite{chandra2001parallel}.

\subsection{System Integration}

The proposed approach, namely \textit{FlowTune}, is implemented in ABC and LSOracle \cite{lso}. Using the I/O interfaces of ABC and LSOracle, \textit{FlowTune} can take as input logic networks in various formats such as Verilog, AIG, and BLIF. The system overview is shown in Figure \ref{fig:system_overview}. On the ABC side, besides integrating \textit{FlowTune} with logic transformations, we also integrated it with two technology mappers, i.e., \texttt{'map'} for standard-cell mapping, and \texttt{'if'} for FPGA mapping. 
Similarly, for MIG evaluation, \textit{FlowTune} was integrated with LSOracle for both logic transformations and a native MIG LUT-mapper, the \texttt{'lut\_map'} command. 
For an accurate evaluation of \textit{FlowTune}, we perform post-technology mapping ASIC evaluation, using ABC + Genus for STA, and end-to-end FPGA evaluation in two different contexts: (i) using ABC + VTR as backend targeting a Stratix IV like architecture, and (ii) using LSOracle + OpenFPGA as back-end, also targeting a Stratix IV like architecture. 

\section{Experimental Results}\label{sec:result}

The experimental results are obtained using a CentOS 7 machine with a 48-core Intel Xeon operating at 2.1 GHz, 8 TB RAM, and 2 TB SSD. We demonstrate the proposed approach using six designs obtained from VTR 8.0 \cite{luu2014vtr}. FlowTune offers high flexibility and effectiveness in optimizing the logic synthesis procedures. Thus, we present results for different scenarios, from technology independent optimization to post mapping results for ASIC, and post-PnR results for FPGA. 
Particularly, in this work, we focus on end-to-end design scenarios to demonstrate the effectiveness at post-PnR stage, including post-PnR assessment of AIGs + VPR flow and MIGs + OpenFPGA flow. Table \ref{tbl:benchmark} summarizes the number of I/O pins, the number of nodes, the logic depth, and a number of latches of the selected benchmarks. In both cases, i.e., when we consider VTR flow with AIG, or MIG-based flow for OpenFPGA, the number of initial nodes are the same. {Note that all synthesis transformations for MIG and AIG flows are all remaining unchanged w.r.t their original implementations.} 

\begin{table}[h]
\centering
\caption{Details of selected VTR benchmarks for evaluating FlowTune. The designs are converted into BLIF format using VTR flow.}
\begin{tabular}{|c|c|c|c|c|}
\hline
\textbf{Design} & \textbf{I/O} & \textbf{Nodes} & \textbf{Latch} & \textbf{Level} \\ \hline
\textit{bfly} & 482/257 & 28910 & 1748 & 97 \\ \hline
\textit{dscg} & 418/257 & 28252 & 1618 & 92 \\ \hline
\textit{fir} & 450/225 & 27704 & 1882 & 94 \\ \hline
\textit{ode} & 275/169 & 16069 & 1316 & 98 \\ \hline
\textit{or1200} & 588/509 & 12833 & 670 & 148 \\ \hline
\textit{syn2} & 450/321 & 30003 & 1512 & 93 \\ \hline
\end{tabular}
\label{tbl:benchmark}
\end{table}

\subsection{Standard Cell Technology Mapping}\label{sec:result:std}

It is known that technology-independent metrics have often miss-correlation with post-technology mapping results \cite{liu2017parallelized}. Yet, many logic synthesis works limit the results and analysis to node count and logic depth, instead of actual post-mapping area and delay. In this work, we aim to show that \textit{FlowTune} can find good solutions post-mapping, and is not limited to DAG size and depth reduction. With this experiment, we show that \textit{FlowTune} is portable and easy to integrate, besides improving the desing post-mapping, which is ubiquitous to make this approach practical.

\begin{figure*}[!htb]
\centering
\begin{minipage}{0.3\textwidth}
  \centering
\includegraphics[width=1\textwidth]{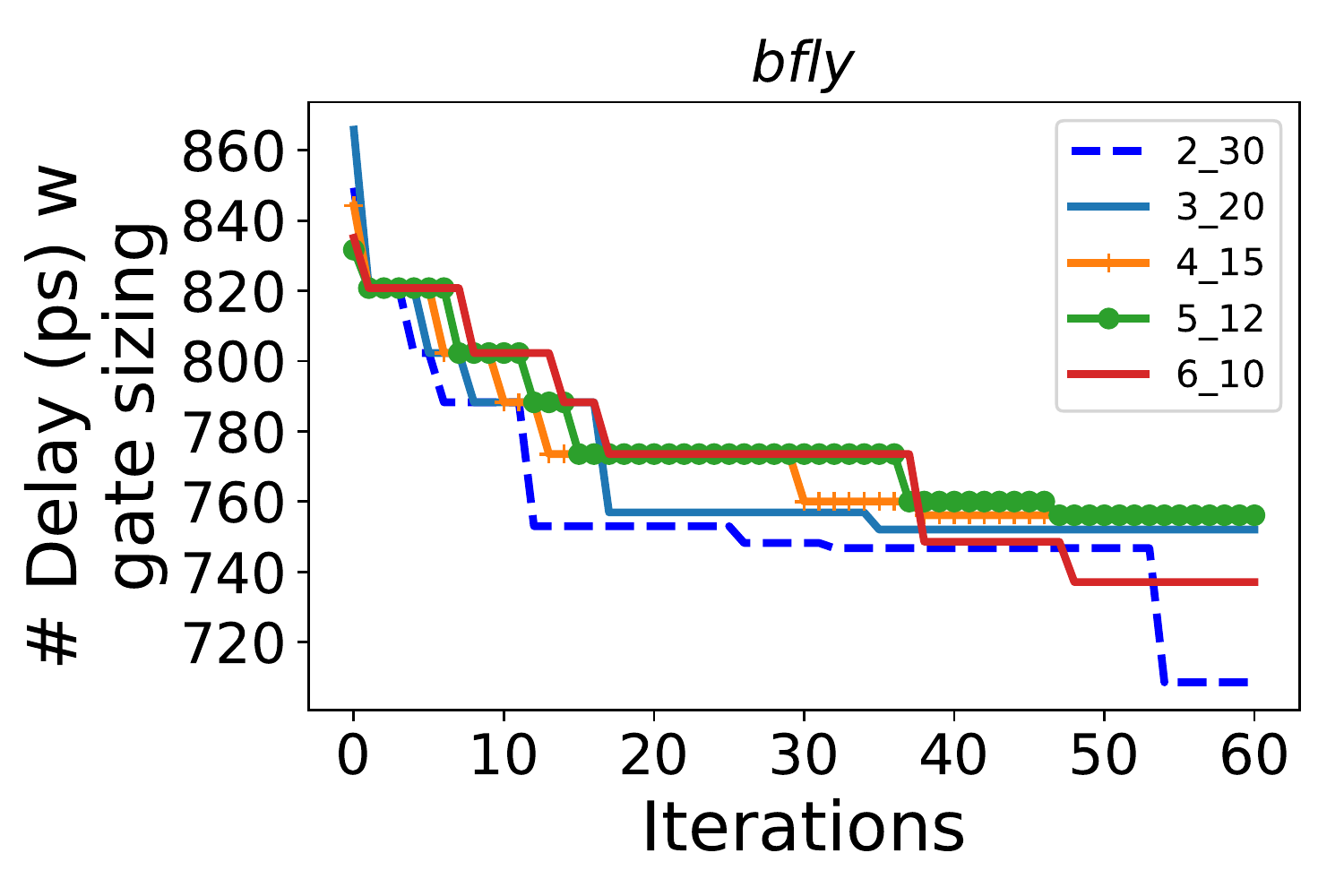}
\end{minipage}
\begin{minipage}{0.3\textwidth}
  \centering
\includegraphics[width=1\textwidth]{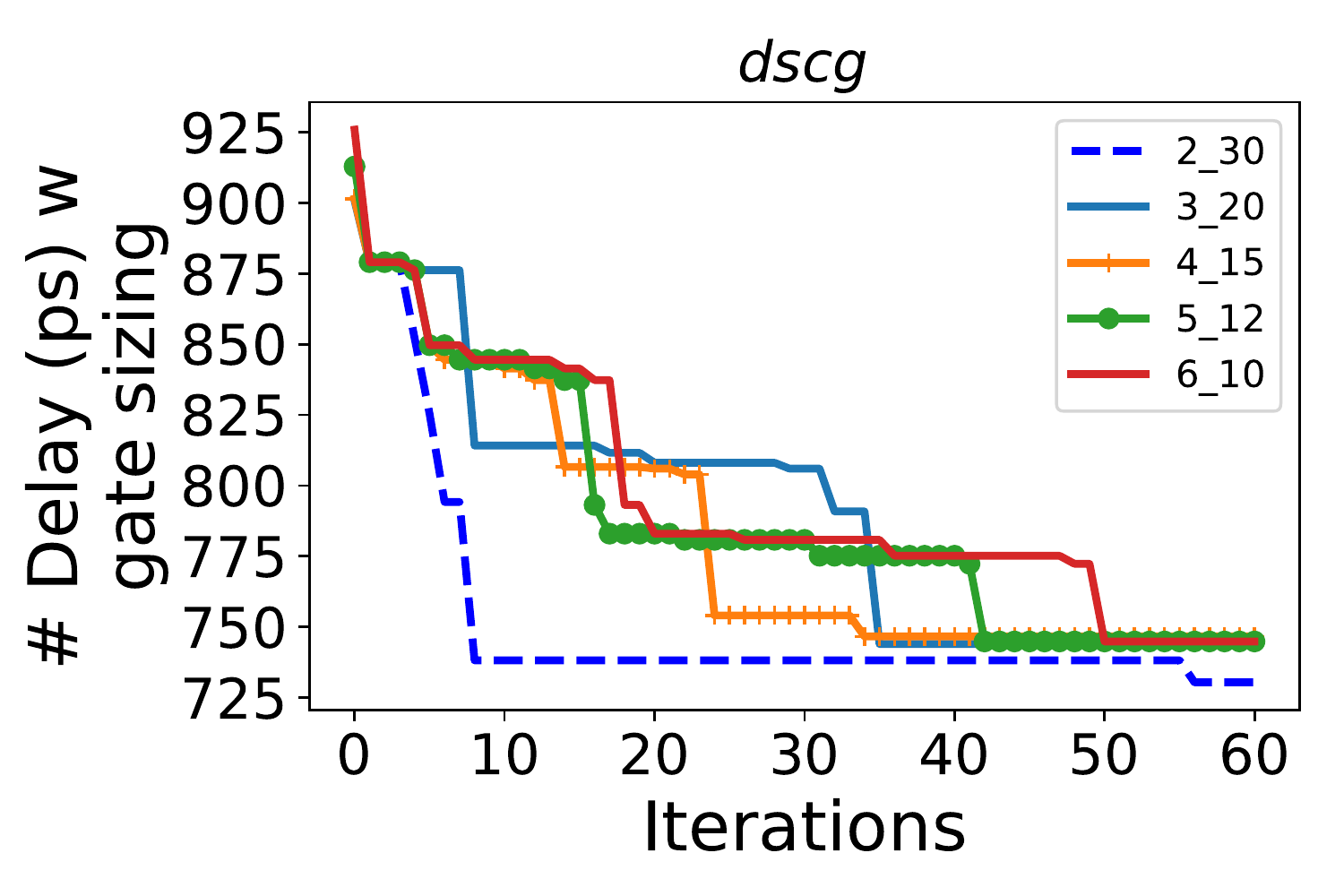}
\end{minipage}
\hspace{0mm}
\begin{minipage}{0.3\textwidth}
  \centering
\includegraphics[width=1\textwidth]{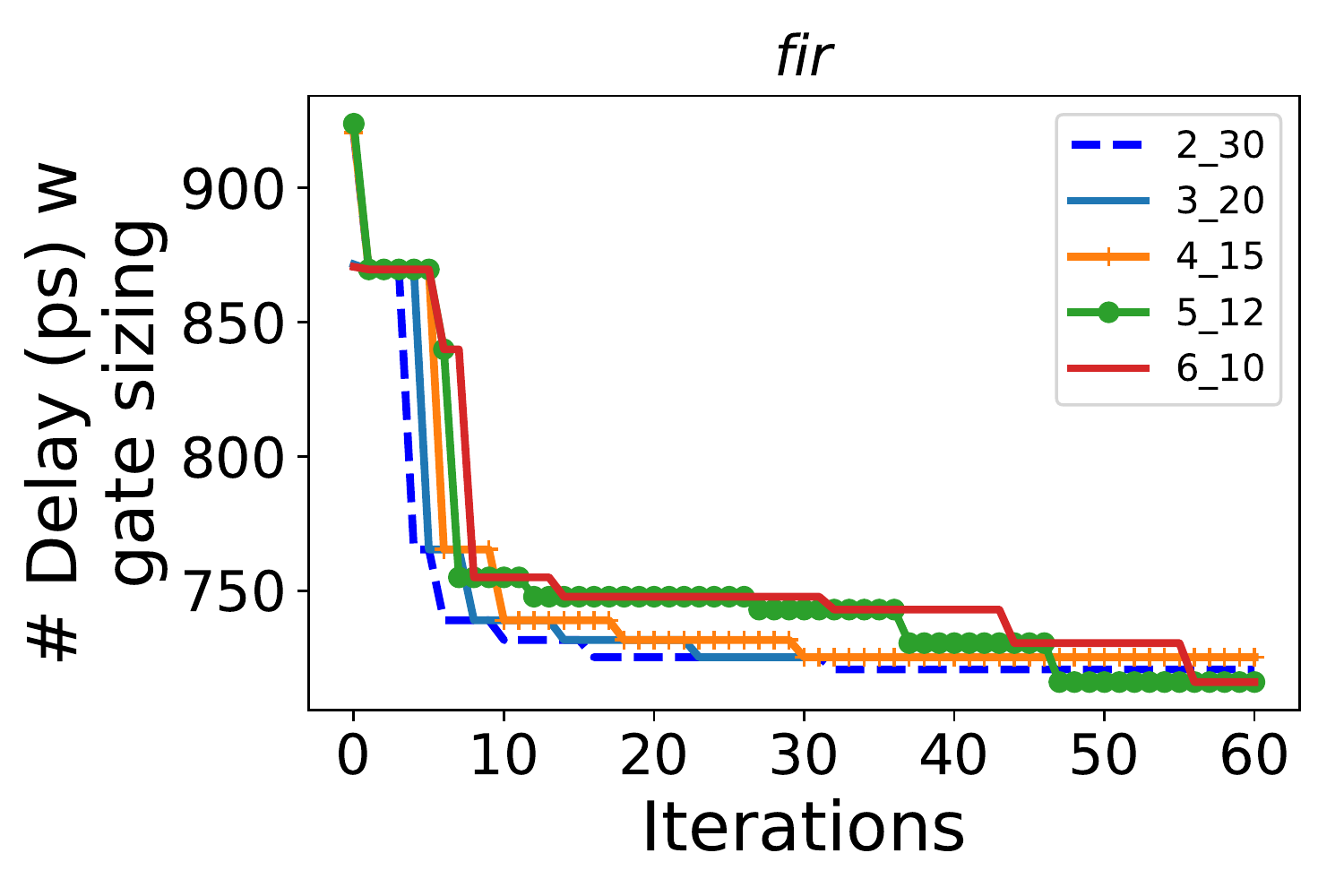}
\end{minipage}
\hspace{0mm}
\begin{minipage}{0.3\textwidth}
  \centering
\includegraphics[width=1\textwidth]{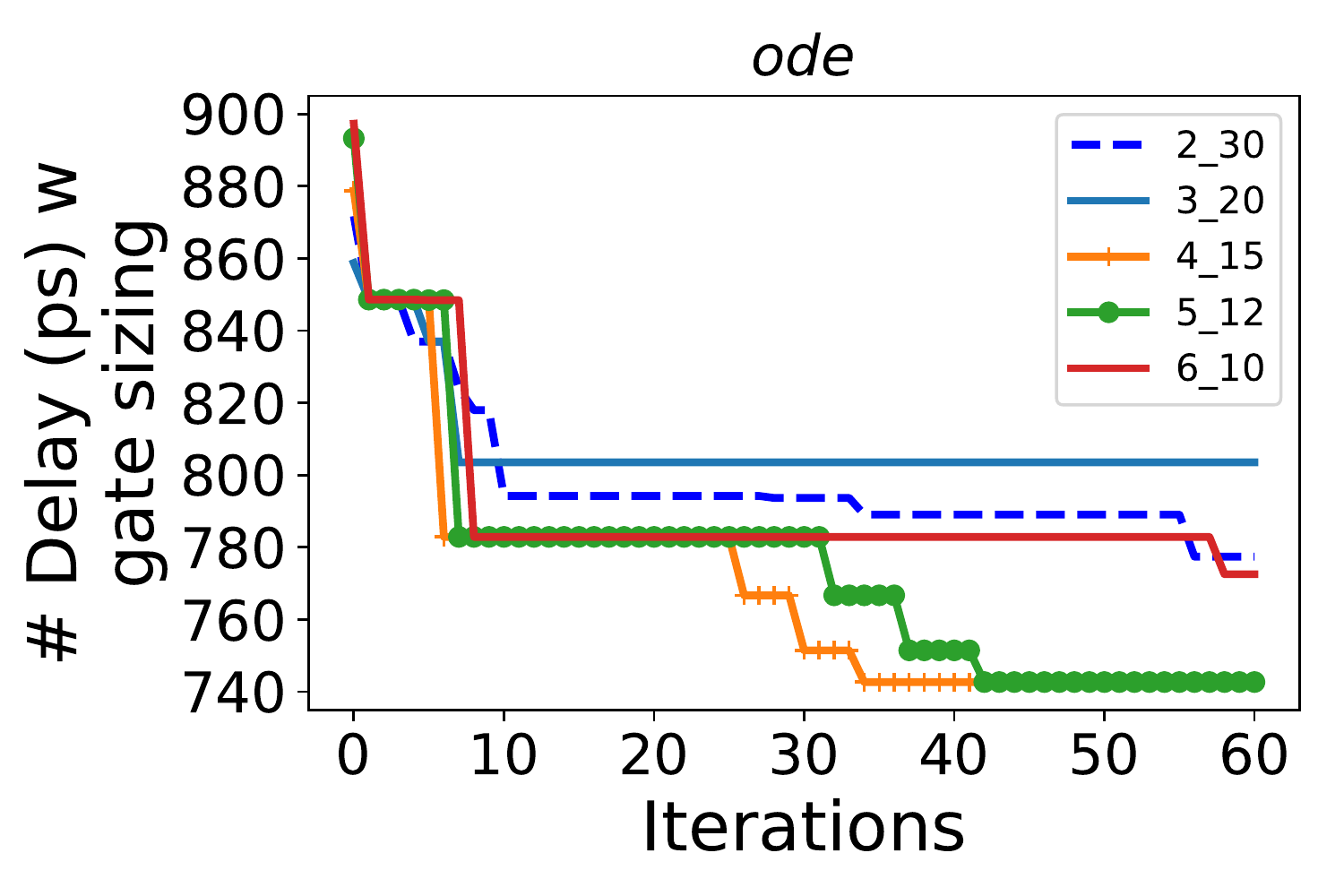}
\end{minipage}
\begin{minipage}{0.3\textwidth}
  \centering
\includegraphics[width=1\textwidth]{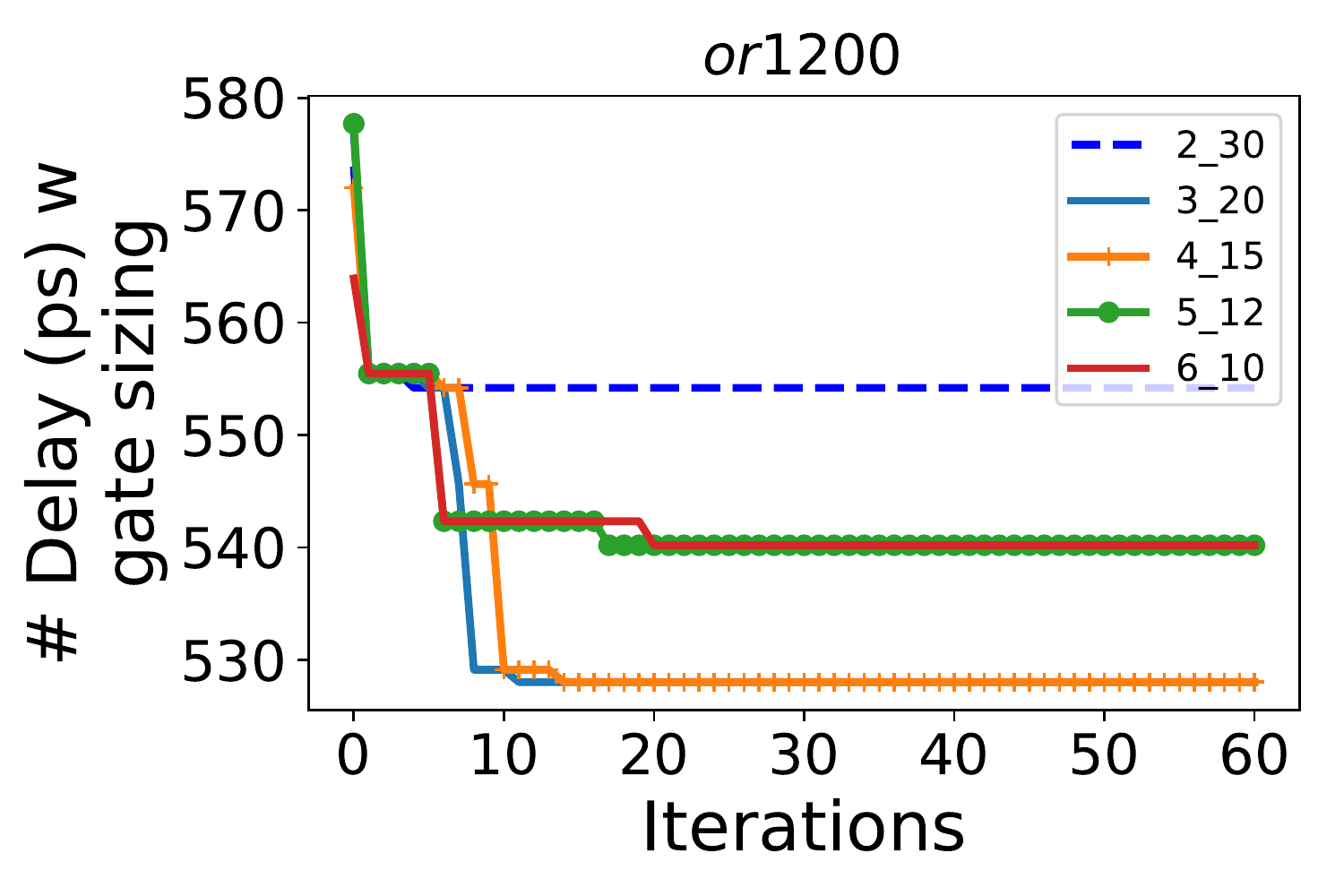}
\end{minipage}
\begin{minipage}{0.3\textwidth}
  \centering
\includegraphics[width=1\textwidth]{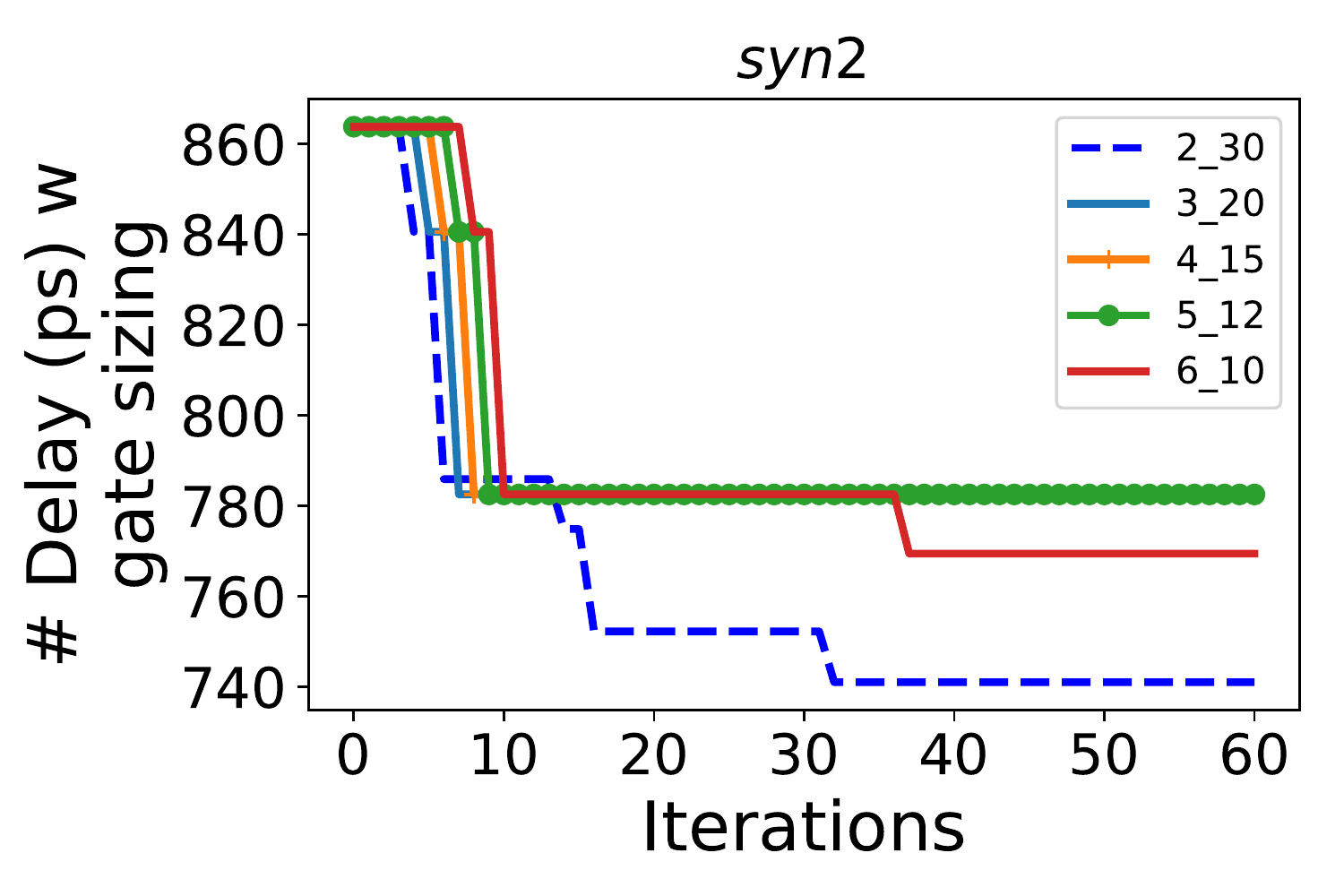}
\end{minipage}
\vspace{-3mm}
\caption{\textit{FTune} in STA delay optimization with gate sizing. QoR results are obtained using Genus with ASAP 7nm library.}
\label{fig:result_std_delay}
\end{figure*}

\begin{figure*}[!htb]
\centering
\begin{minipage}{0.3\textwidth}
  \centering
\includegraphics[width=1\textwidth]{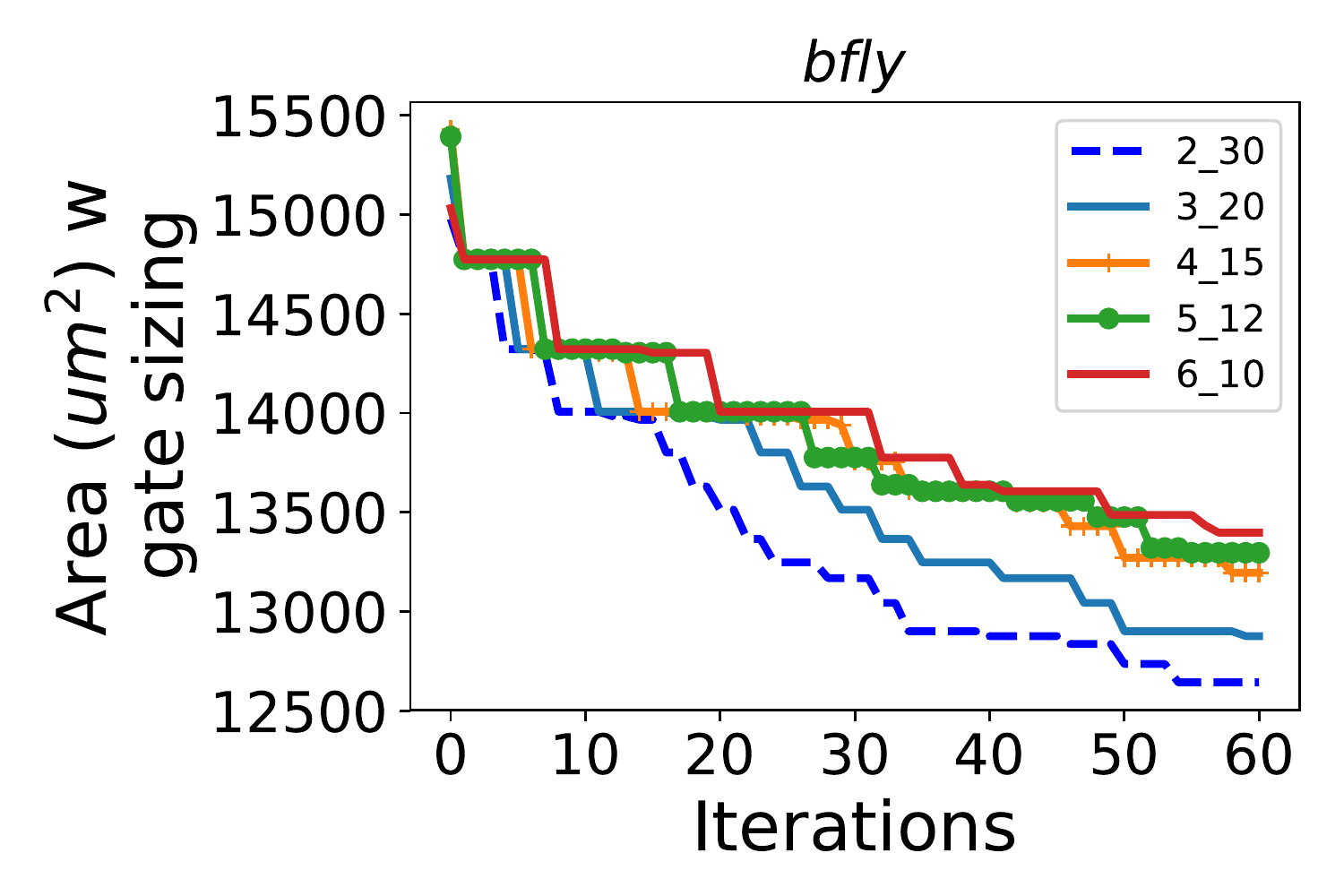}
\end{minipage}
\begin{minipage}{0.3\textwidth}
  \centering
\includegraphics[width=1\textwidth]{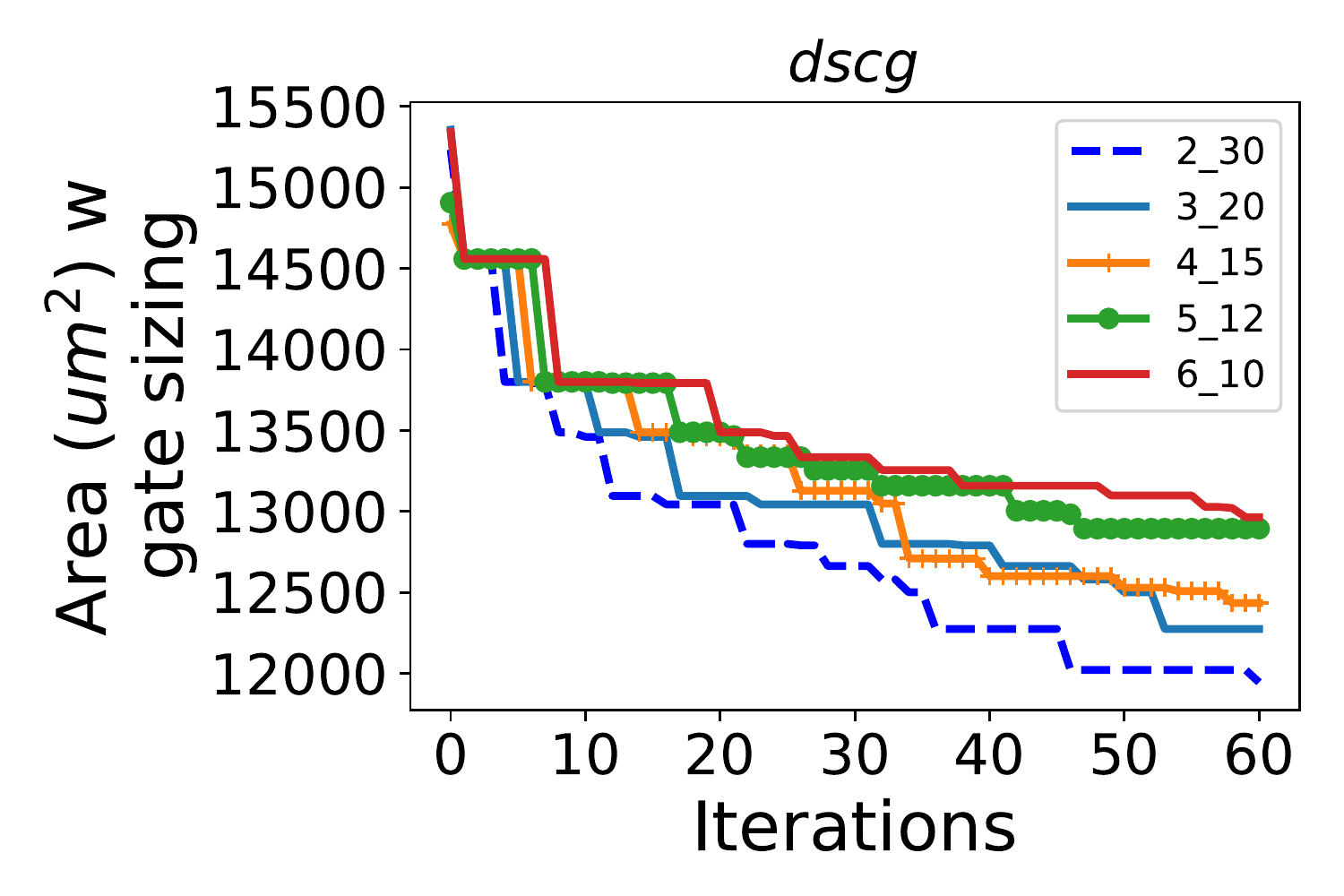}
\end{minipage}
\hspace{0mm}
\begin{minipage}{0.3\textwidth}
  \centering
\includegraphics[width=1\textwidth]{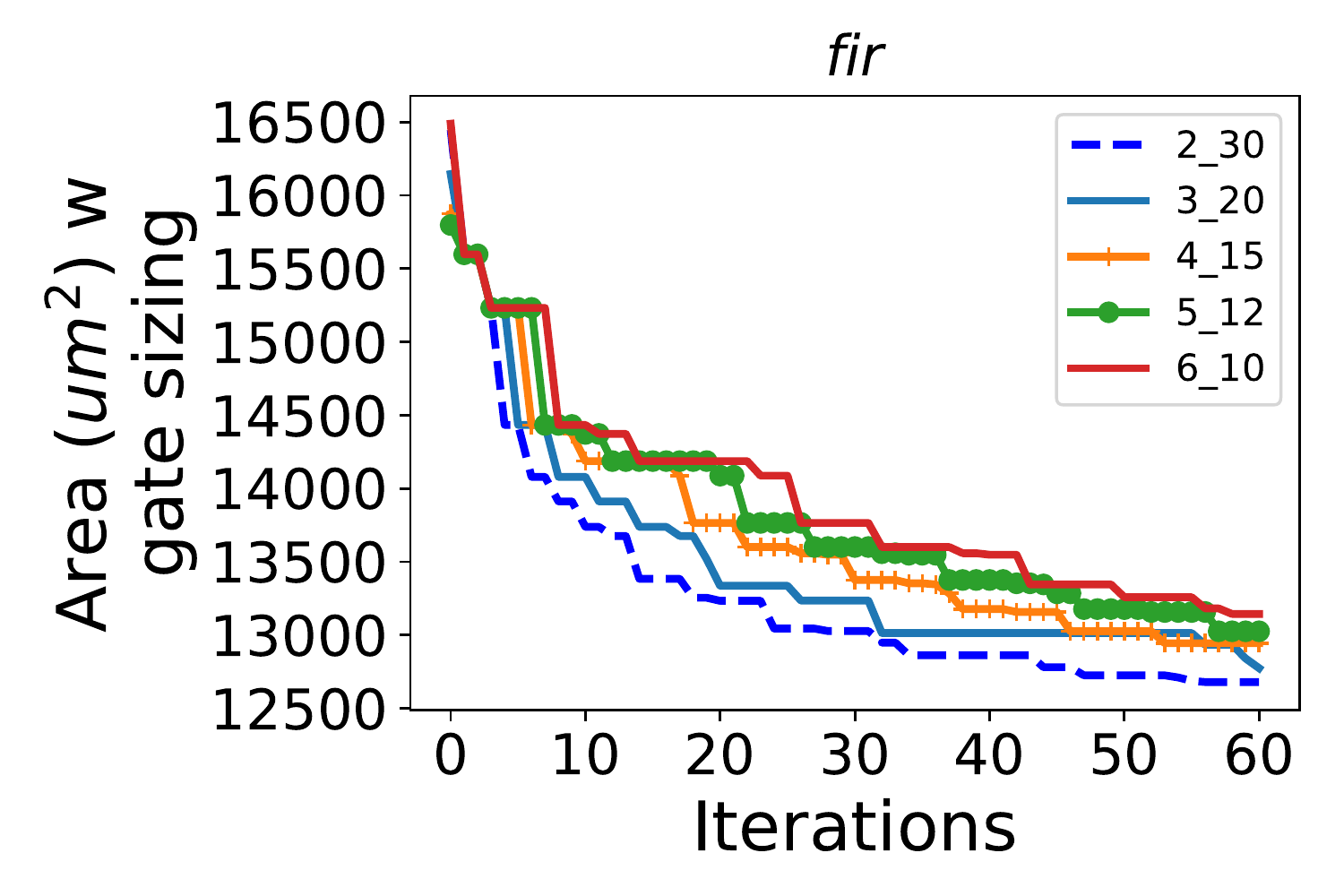}
\end{minipage}
\hspace{0mm}
\begin{minipage}{0.3\textwidth}
  \centering
\includegraphics[width=1\textwidth]{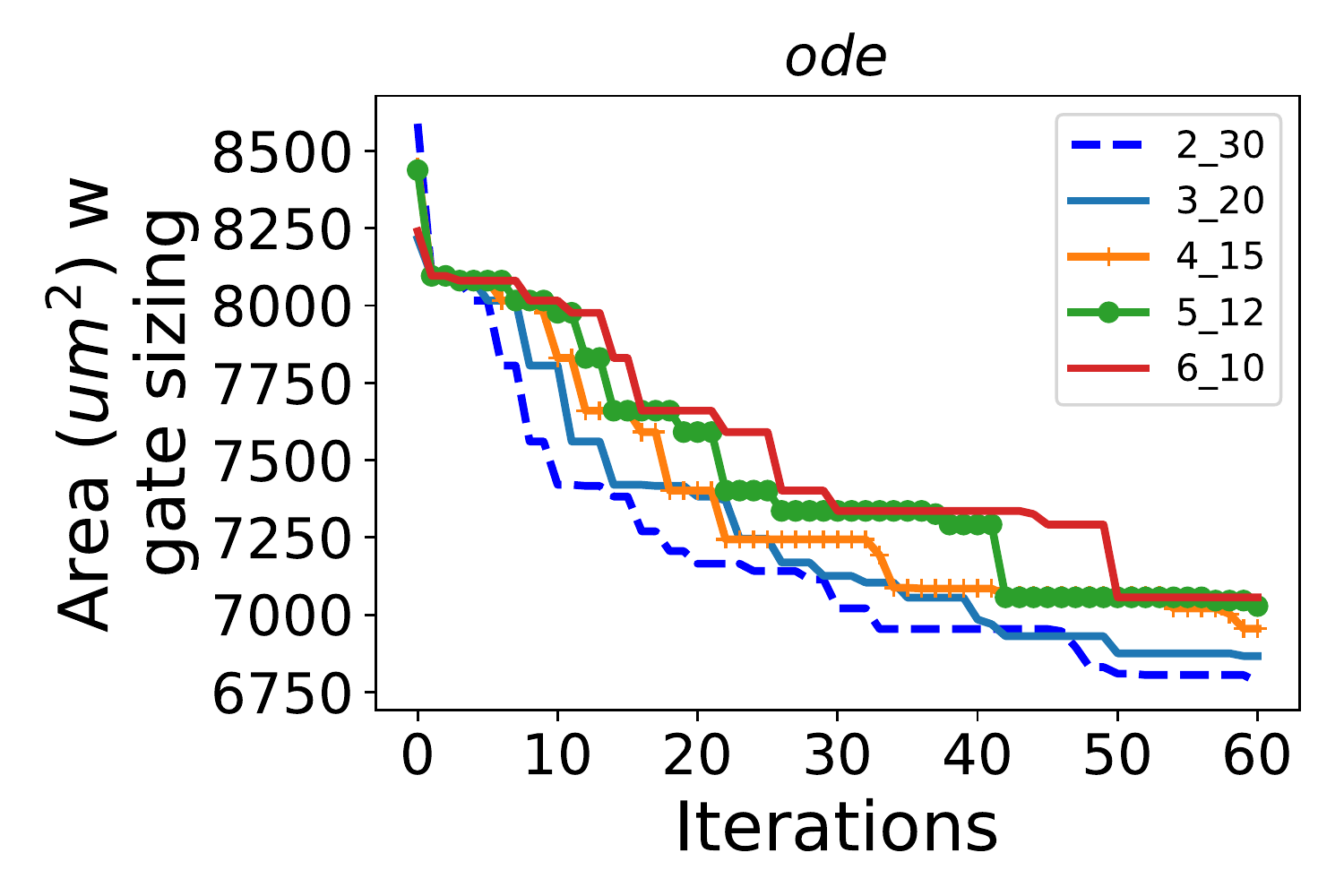}
\end{minipage}
\begin{minipage}{0.3\textwidth}
  \centering
\includegraphics[width=1\textwidth]{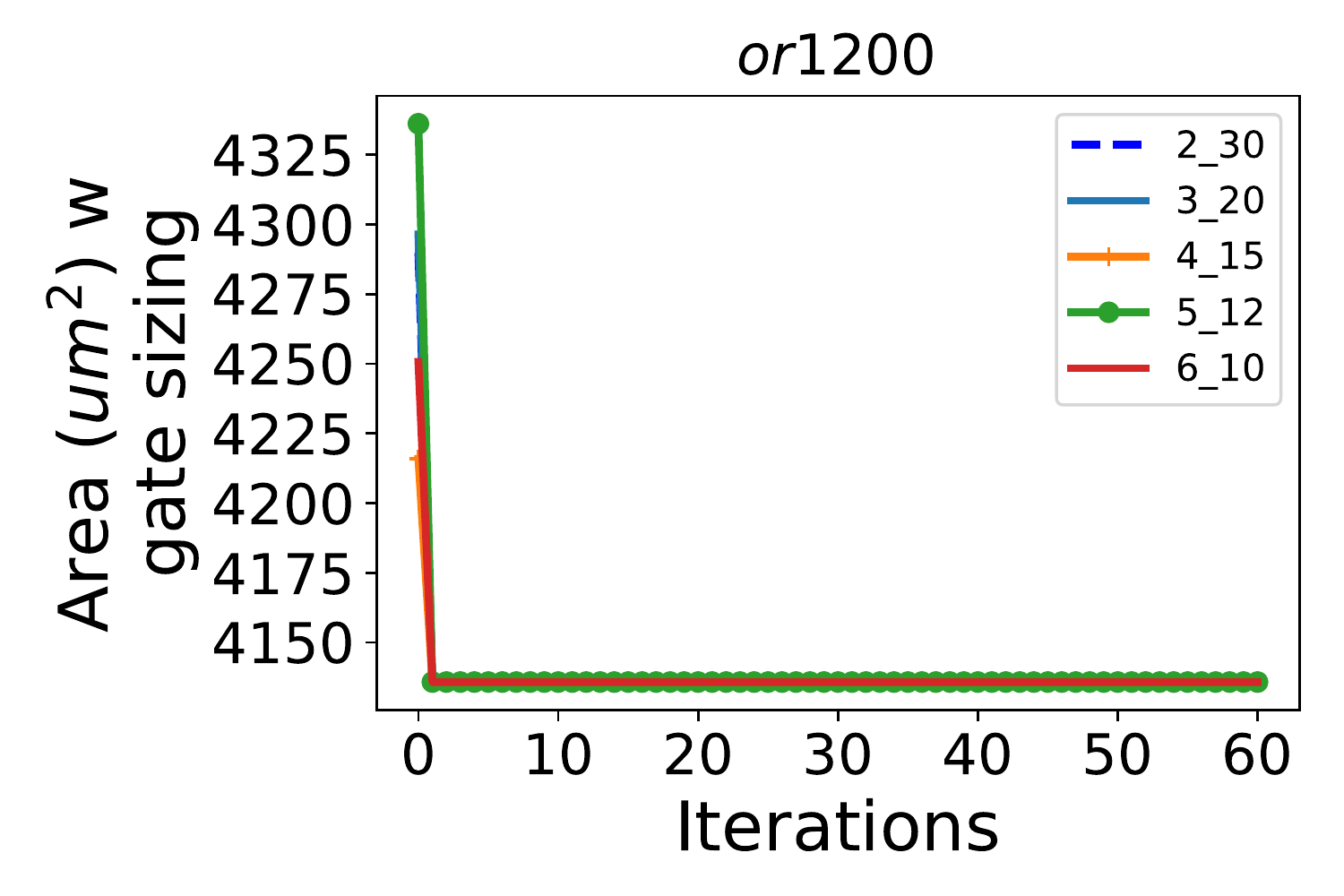}
\end{minipage}
\begin{minipage}{0.3\textwidth}
  \centering
\includegraphics[width=1\textwidth]{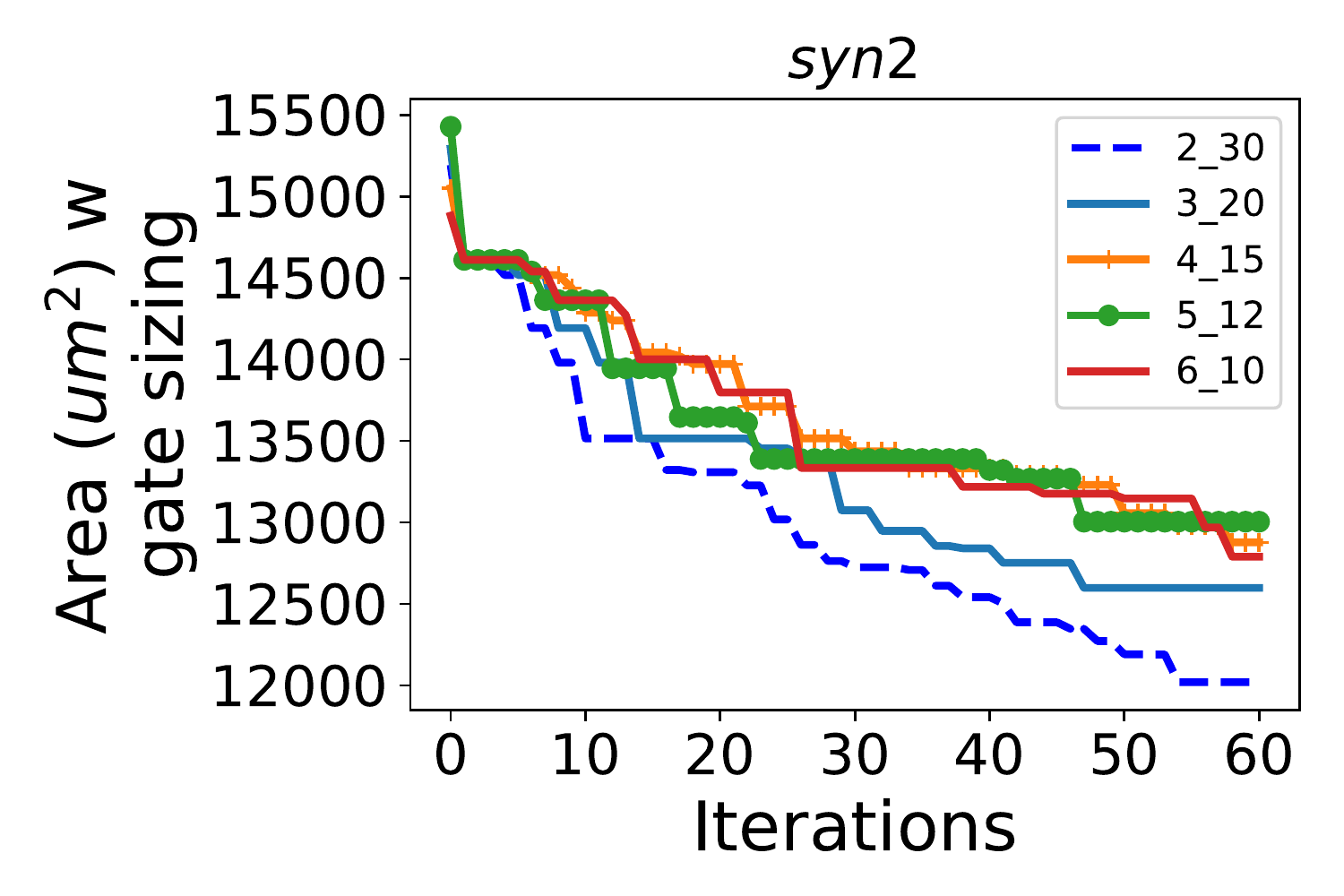}
\end{minipage}
\vspace{-3mm}
\caption{\textit{FTune} in area optimization using gate sizing. QoR results are obtained using Genus with ASAP 7nm library.}
\label{fig:result_std_area}
\end{figure*}

\textbf{STD technology mapping optimization with FlowTune}: We aim to optimize the technology-mapping QoR, evaluated by Cadence Genus {with gate sizing}, using the ASAP 7nm library, while targeting \textbf{a)} STA delay optimization (Figure \ref{fig:result_std_delay}); and \textbf{b)} area optimization (Figure \ref{fig:result_std_area}). QoR results are collected with Genus by importing the mapped Verilog from ABC. To the best of our knowledge, this is the first work that addresses synthesis flow tuning for STA-aware technology mapping. 

We can see that \textit{FlowTune} effectively explores the design space by finding better synthesis flows for both area and post-STA delay optimization. However, \textit{FlowTune} is not able to find any better synthesis flow after the initialization for design \texttt{or1200}. This is similar to what we observed during the FPGA experiments for this same design \ref{fig:vpr_area}. We believe the ABC synthesis flow design space of \texttt{or1200} is very limited. While for delay different \textit{FlowTune} setups perform better in different designs, for area a setup ($s:m$ = 2:30) performs consistently better than others. Thus, while targeting area optimization, \textit{FlowTune} can be set with $s:m$ = 2:30. {The runtime cost of FlowTune of all benchmarks listed in Table \ref{tbl:benchmark} vary from 433 seconds to 1104 seconds, with average 892 seconds runtime cost, as depicted in Table IV, for AIG optimization. Note that the runtime overhead of FlowTune is the same regardless if used with AIGs or MIGs. Thus, the overhead should be constant regardless the DAG, and runtime differences are due to the optimization for the different DAGs. Also, for larger designs partitioning could be used to run FlowTune in parallel and reduce the runtime.  }

\begin{table}[t]

\centering
 {\caption{FlowTune runtime for synthesis exploration for AIGs.}}
\resizebox{\columnwidth}{!}{%
\begin{tabular}{|l|clllll|}
\hline
\multirow{2}{*}{} &
  \multicolumn{6}{c|}{\textbf{Design}} \\ \cline{2-7} 
 &
  \multicolumn{1}{c|}{\textit{bfly}} &
  \multicolumn{1}{c|}{\textit{dscg}} &
  \multicolumn{1}{c|}{\textit{fir}} &
  \multicolumn{1}{c|}{\textit{ode}} &
  \multicolumn{1}{c|}{\textit{or1200}} &
  \multicolumn{1}{c|}{\textit{syn2}} \\ \hline
\textbf{Runtime(s)} &
  \multicolumn{1}{l|}{1,101.94} &
  \multicolumn{1}{l|}{1,052.88} &
  \multicolumn{1}{l|}{1,049.94} &
  \multicolumn{1}{l|}{611.96} &
  \multicolumn{1}{l|}{433.93} &
  1,104.80 \\ \hline
\end{tabular}%
}
\end{table}


\subsection{End-to-End Integration and Evaluation for FPGA Design}

One of the greatest challenges of leveraging ML techniques in design flow optimization is to demonstrate the optimizations can be fully realized in an end-to-end design process, e.g., evaluating the QoR performance at the post-routing stage. Therefore, in this section, we evaluate \textit{FlowTune} in two different end-to-end FPGA design frameworks, VTR 8.0 \cite{luu2011vpr} and OpenFPGA \cite{tang2019openfpga}, where our framework is integrated at the logic synthesis stage in both frameworks, for AIG and MIG. With this experiment, we aim to show the flexibility and portability of \textit{FlowTune}, along with its capabilities to improve QoR. Thus, note that our goal is not to compare VTR and OpenFPGA, or AIGs and MIGs.

\noindent
\textbf{Evaluation metrics} -- To comprehensively evaluate the FPGA implementation performance, a list of QoR metrics that cover the area (utilization) and timing are selected. Specifically, our experimental results are conducted on measuring the post-routing (a) \textit{total wirelength} (WL), (b) \textit{total area} (Area) including logic area and routing area, (c) critical path delay, and (d) total negative slack (TNS). 

\noindent
\textbf{Evaluation baselines} -- The baseline results for VTR 8.0 are generated with the default settings collected from the VTR repository. Regarding the OpenFPGA framework, we integrate \textit{FlowTune} in LSOracle, a state-of-the-art logic optimizer that handles MIG and AIGs. We focus on the MIG manipulation, and compare \textit{FlowTune} against a high-effort MIG flow in LSOracle. OpenFPGA is used with its default settings. { Besides the use of well stablished flows for AIG and MIG optimization, we have considered the use of random sampling to see how our approach compares to it. Given a timing budget of 30 minutes (which exceeds our greatest runtime), random sampling could explore $\approx$ 70 flows for \textit{bfly} on ABC, and all these flows presented worst QoR than our framework. Thus, we consider random sampling to not be a strong baseline, and adopt well-stablished design flows to be our baseline}.
Both experiments target a Stratix IV-like FPGA architecture, which is common adopted modern architecture for FPGA works \cite{gore2021scalable, luu2011vpr}. In this context, each logic block is composed of 10 fracturable LUTs, and 200 routing tracks (channels). 

\noindent
\textbf{Experimental setup for VTR 8.0 as backend} -- The experimental results conducted with VTR 8.0 as complete design flow involves the complete design flow from ODIN synthesis, with behavior Verilog HDL as inputs. The rest of the design flow includes logic synthesis and LUT-based technology mapping using ABC \cite{mishchenko2010abc}, in which \textit{FlowTune} is plugged-in to optimize the design with automatically explored design flows. The output design will then be placed and routed w.r.t to the given FPGA architecture using VPR default settings. The results included in this section are all collected at the post-routing stage. 

\noindent
\textbf{{Experimental setup for OpenFPGA as backend}} -- The OpenFPGA design flow also includes the complete design flow from ODIN synthesis, with behavior Verilog HDL as inputs. Then it uses LSOracle to read an input BLIF into a MIG, and perform logic optimization. \textit{FlowTune} is integrated with LSOracle to autonomously generate a MIG-based flow. LSOracle is then used for LUT mapping, and the output is dumped into a BLIF file. The BLIF is then used as input to OpenFPGA with default settings, and all the results are collected post-routing. {Note that LSOracle synthesis framework is MIG-based logic synthesis engine, where the transformations are all based on MIG. These experiments aim to demonstrate the DAG-based synthesis domain-specific knowledge extracted from AIG optimization is transferable to MIG as well.}

To show the flexibility of our approach, we present results for two different end-to-end FPGA design evaluations, i.e., FlowTune in VTR 8.0 and FlowTune in LSOracle + OpenFPGA. Note that we do not intend to compare the differences between logic synthesis data structures (i.e., AIG and MIG), and do not intend to compare the performance across different backend frameworks (VTR and OpenFPGA). In summary, our goal with these experiments is to show that FlowTune can be easily integrated and verify its effectiveness in different scenarios.

\begin{figure*}[t]
\centering
\begin{minipage}{0.37\textwidth}
  \centering
\includegraphics[width=1\textwidth]{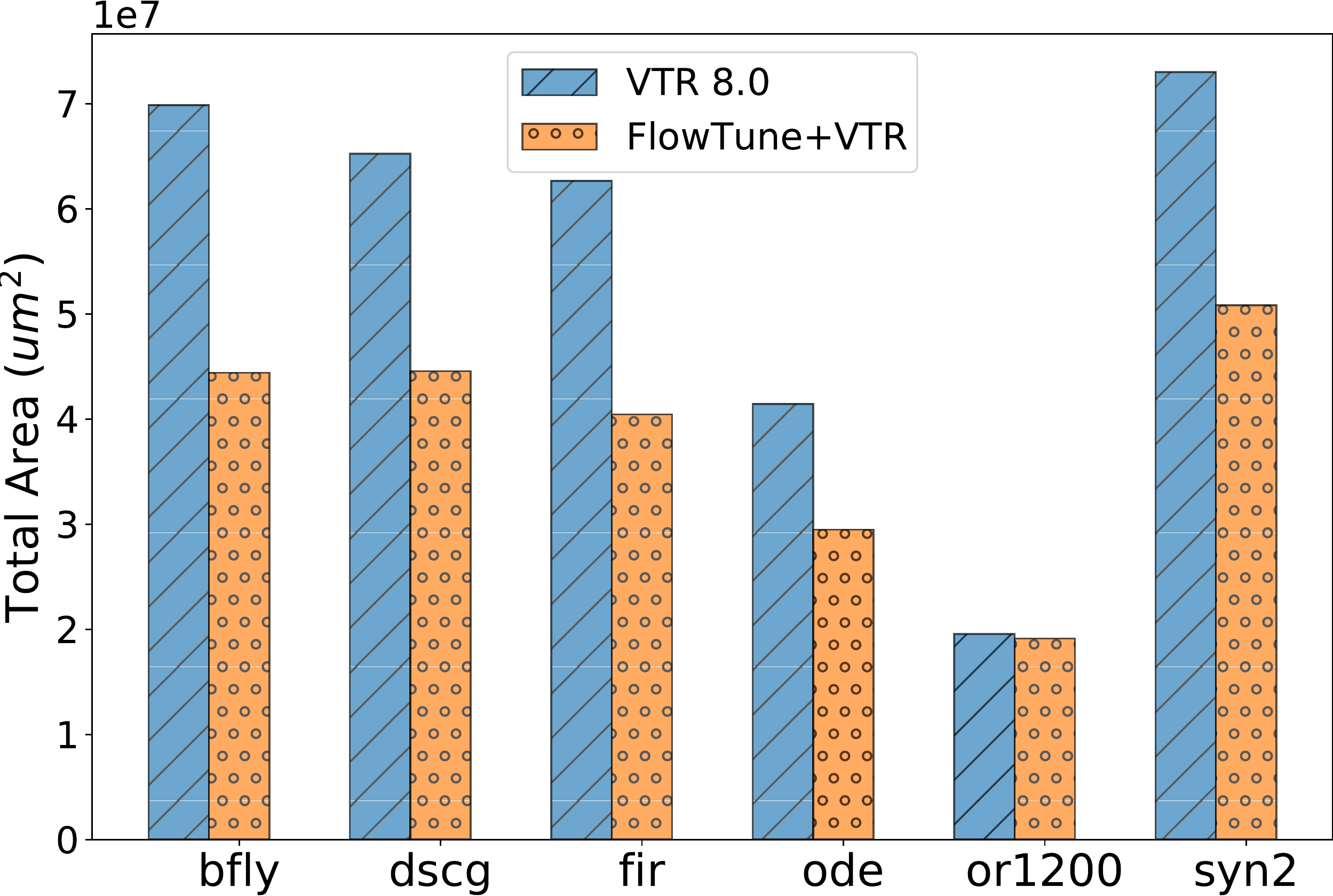}
\subcaption{Total logic and routing area.}\label{fig:vpr_area}
\end{minipage}
\hspace{5mm}
\begin{minipage}{0.37\textwidth}
  \centering
\includegraphics[width=1\textwidth]{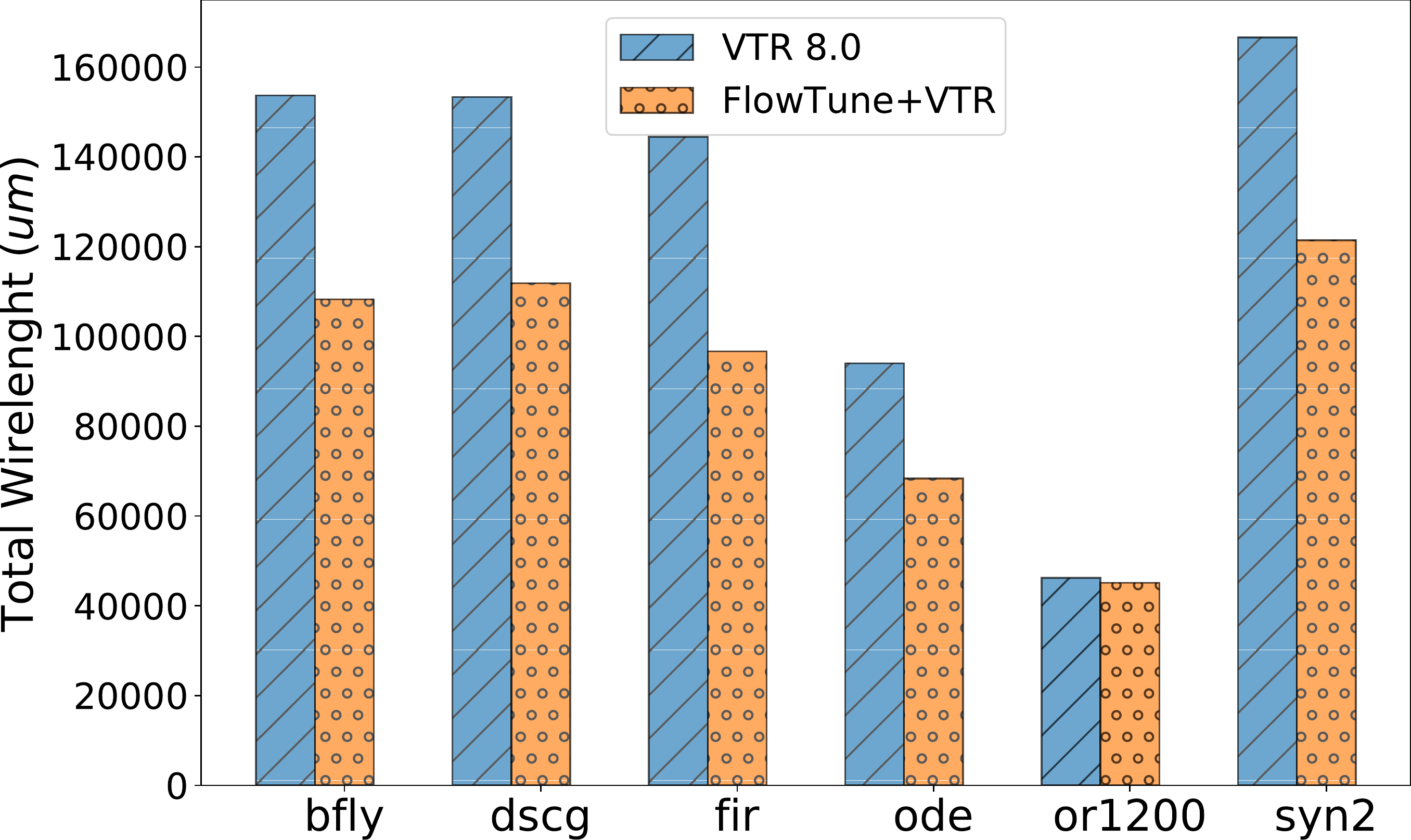}
\subcaption{Total routing wirelength.}\label{fig:vpr_wl}
\end{minipage}
\\
\hspace{0mm}
\begin{minipage}{0.37\textwidth}
  \centering
\includegraphics[width=1\textwidth]{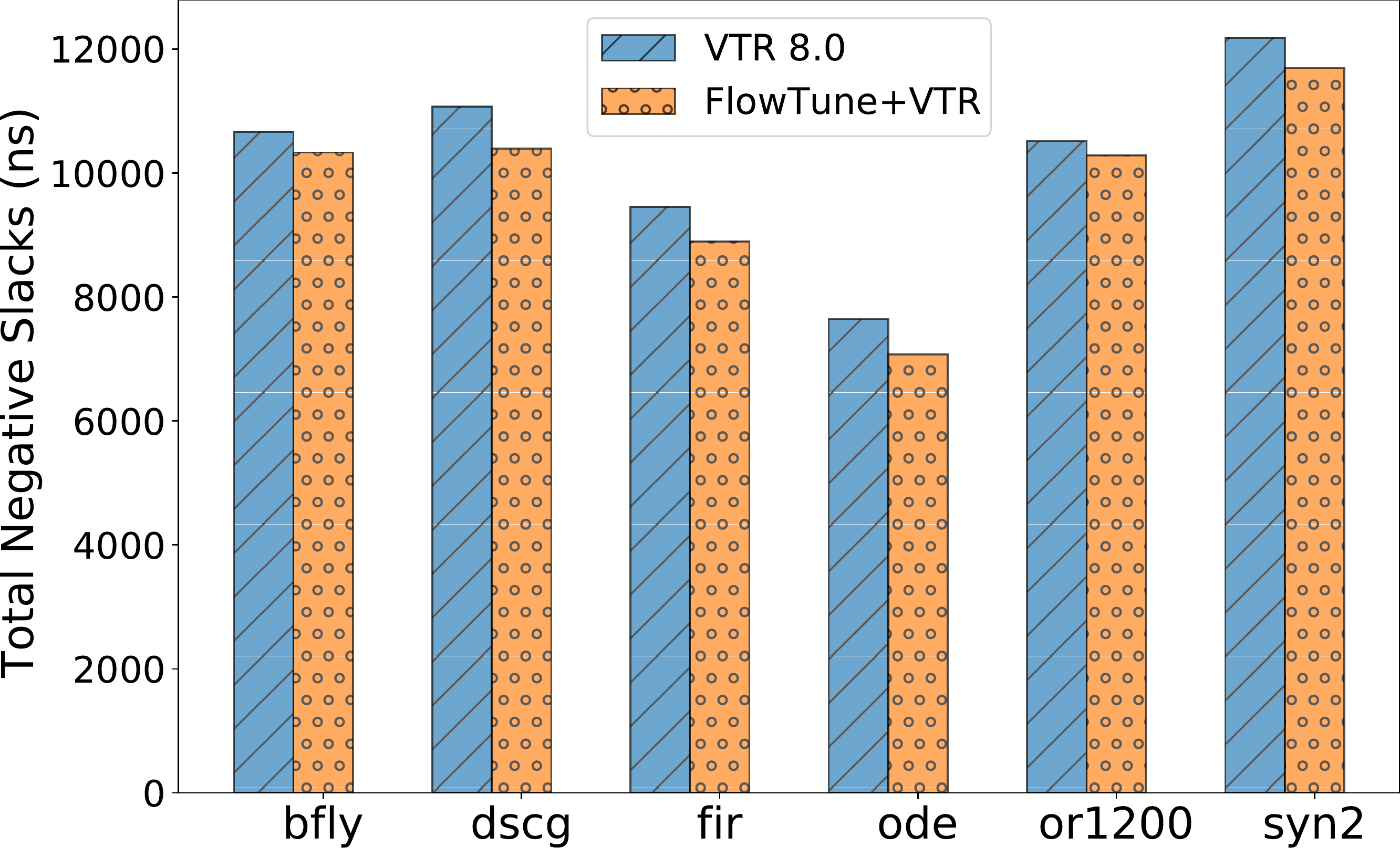}
\subcaption{Post-PnR total negative slacks (TNS).}\label{fig:vpr_tns}
\end{minipage}
\hspace{5mm}
\begin{minipage}{0.37\textwidth}
  \centering
\includegraphics[width=1\textwidth]{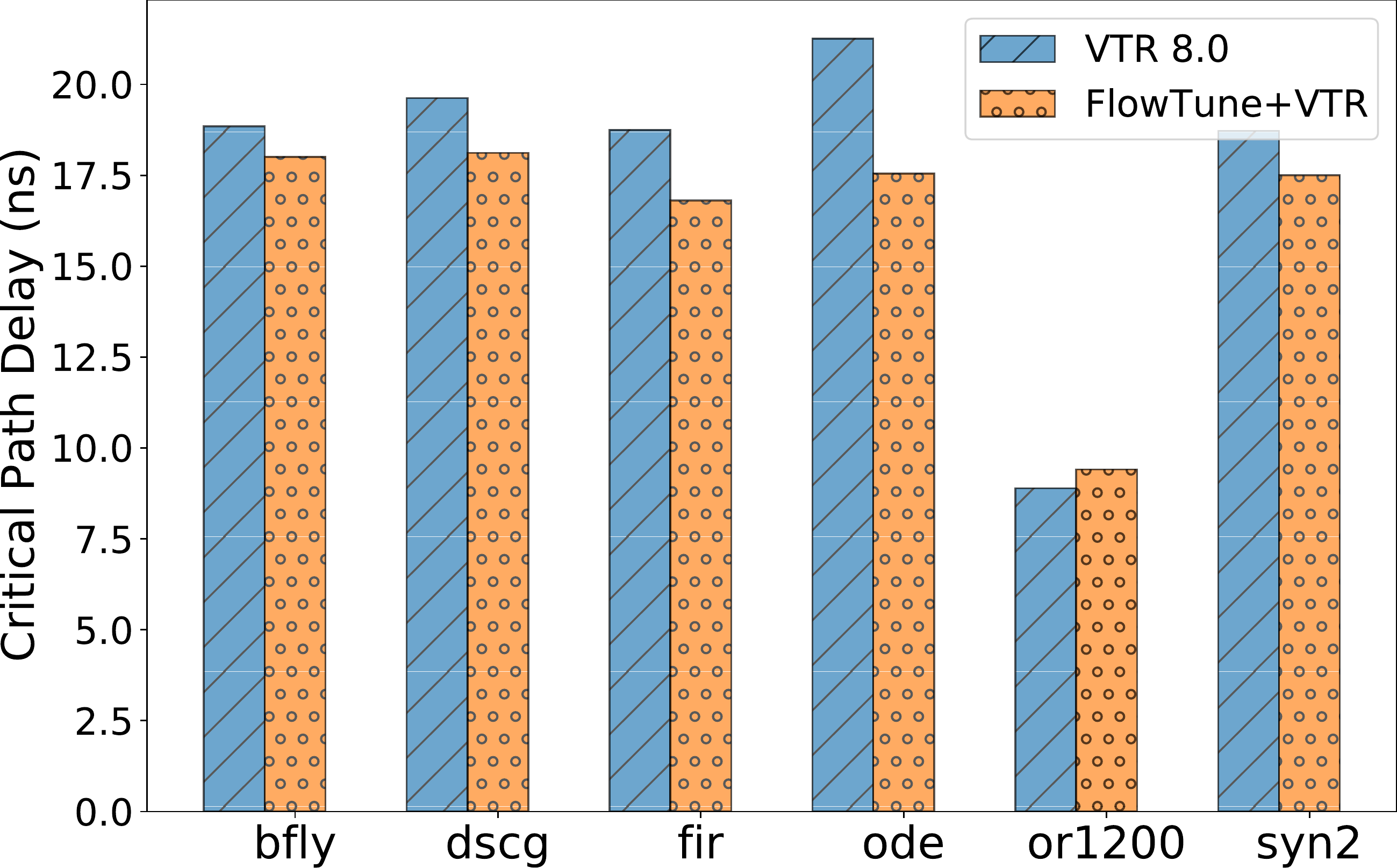}
\subcaption{ Post-PnR critical path delay.}\label{fig:vpr_delay}
\end{minipage}
\caption{ Post-routing evaluation with VTR (VPR PnR) as backend using six benchmarks collected from \cite{luu2014vtr}, with default VTR 8.0 flow as baseline where logic synthesis is conducted on \textbf{AIG logic optimization}. The collected results include (a) total area including logic and routing area, (b) total routing wire length, (c) post-PnR total negative slacks (TNS), and (d) post-PnR critical path delay.}
\label{fig:result_vpr}
\end{figure*}

\begin{figure*}[t]
\centering
\begin{minipage}{0.37\textwidth}
  \centering
\includegraphics[width=1\textwidth]{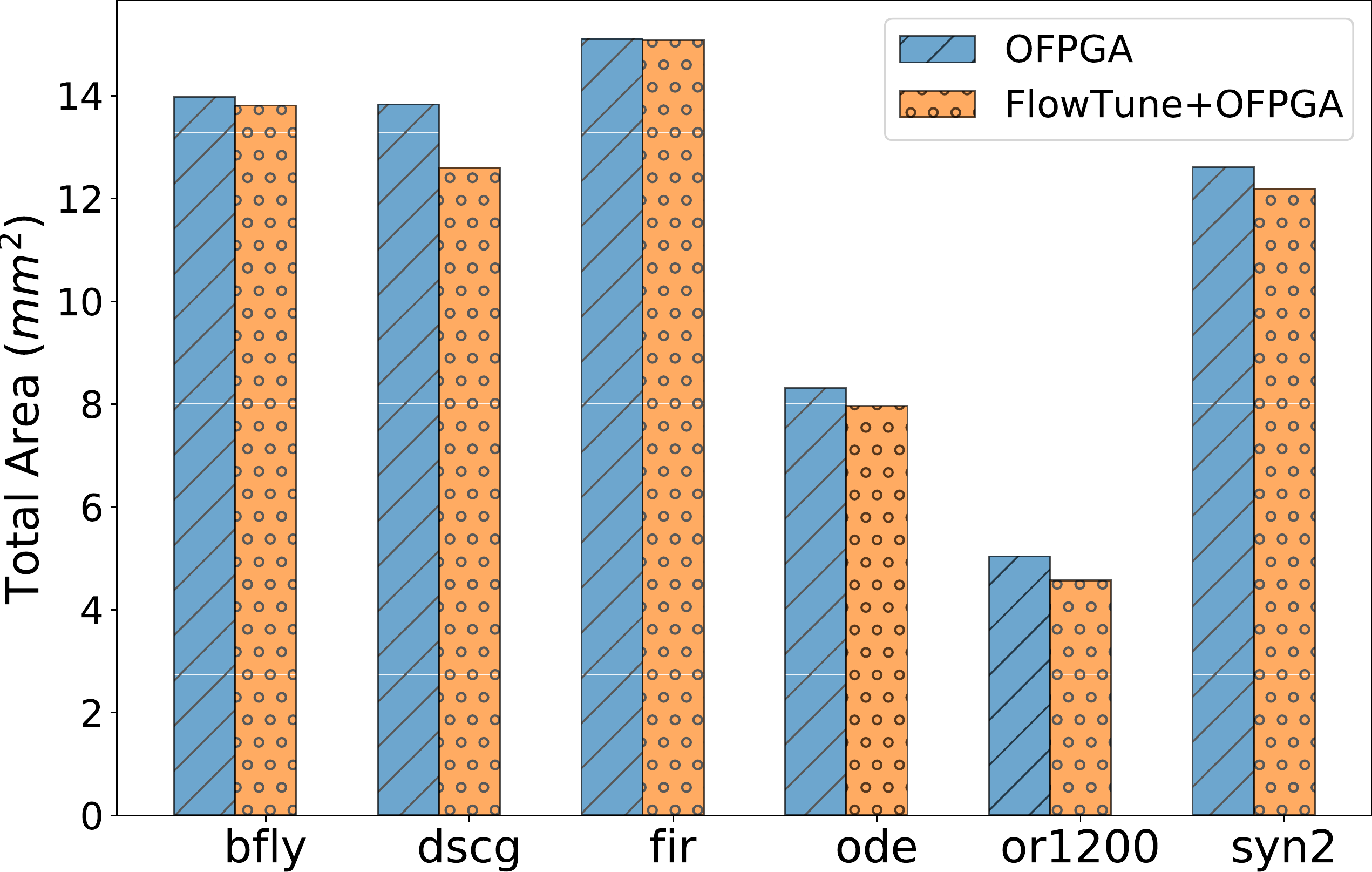}
\subcaption{Total logic and routing area.}\label{fig:ofpga_area}
\end{minipage}
\hspace{5mm}
\begin{minipage}{0.37\textwidth}
  \centering
\includegraphics[width=1\textwidth]{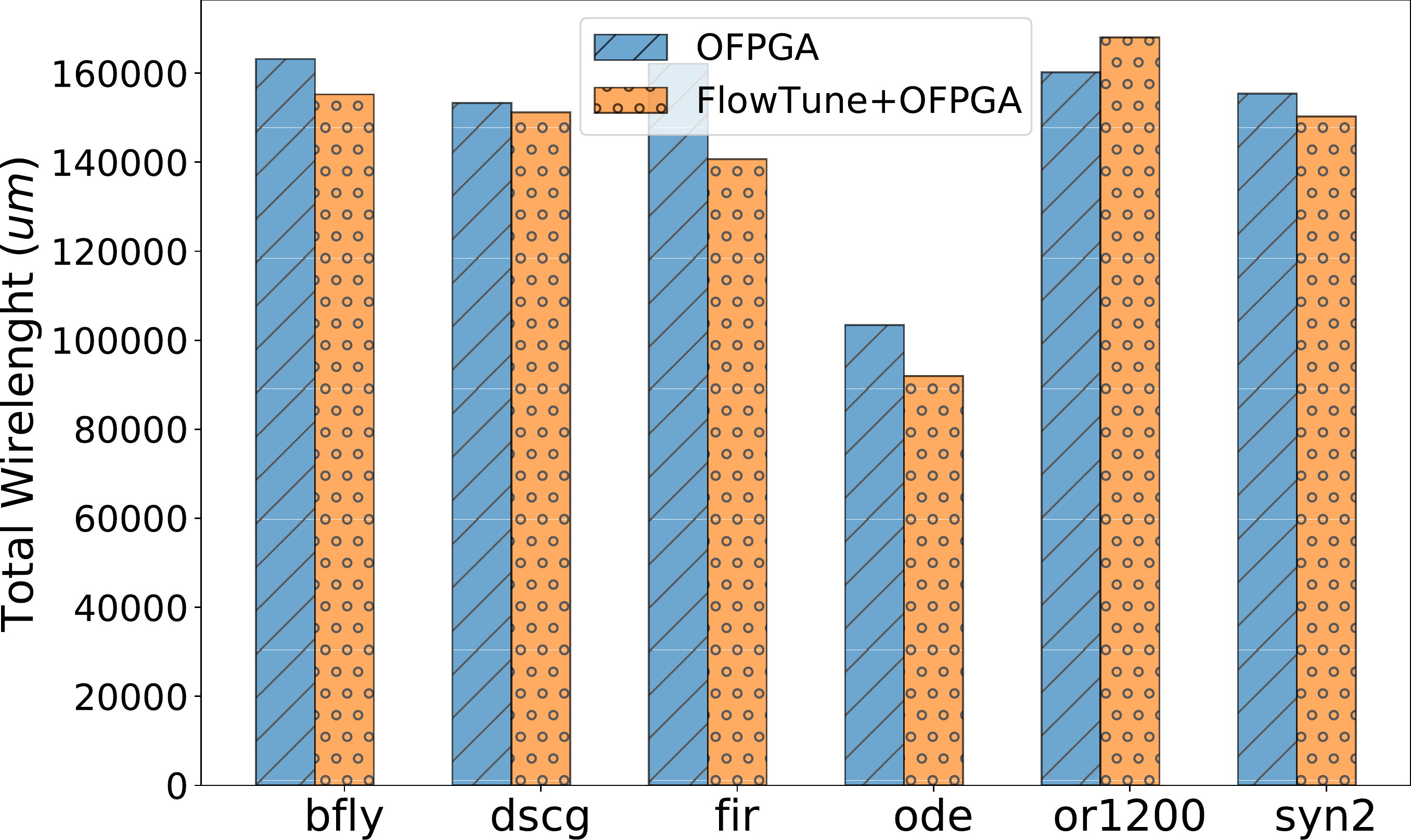}
\subcaption{Total routing wirelength.}\label{fig:ofpga_wl}
\end{minipage}
\\
\hspace{0mm}
\begin{minipage}{0.37\textwidth}
  \centering
\includegraphics[width=1\textwidth]{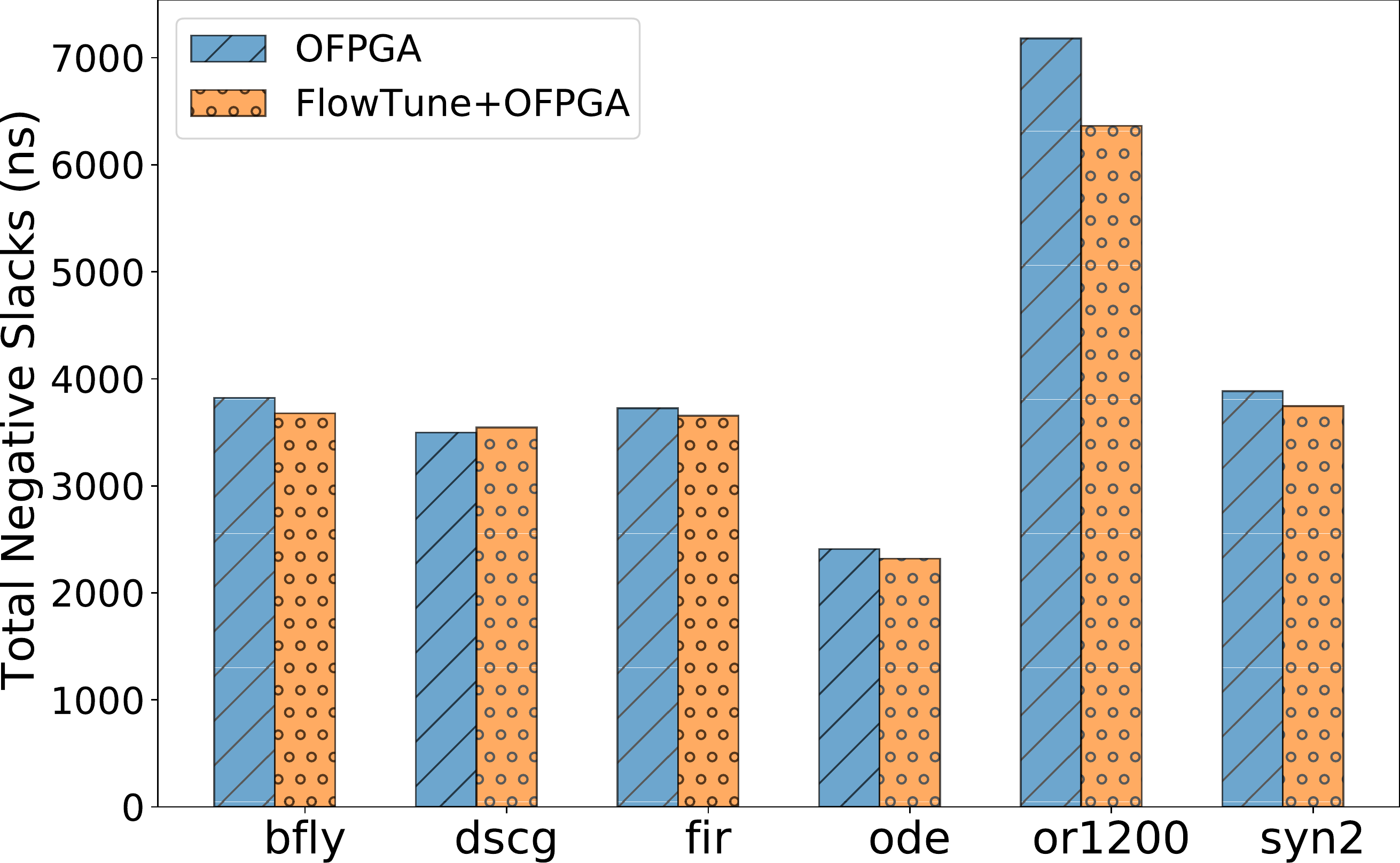}
\subcaption{Post-PnR total negative slacks (TNS).}\label{fig:ofpga_tns}
\end{minipage}
\hspace{5mm}
\begin{minipage}{0.37\textwidth}
  \centering
\includegraphics[width=1\textwidth]{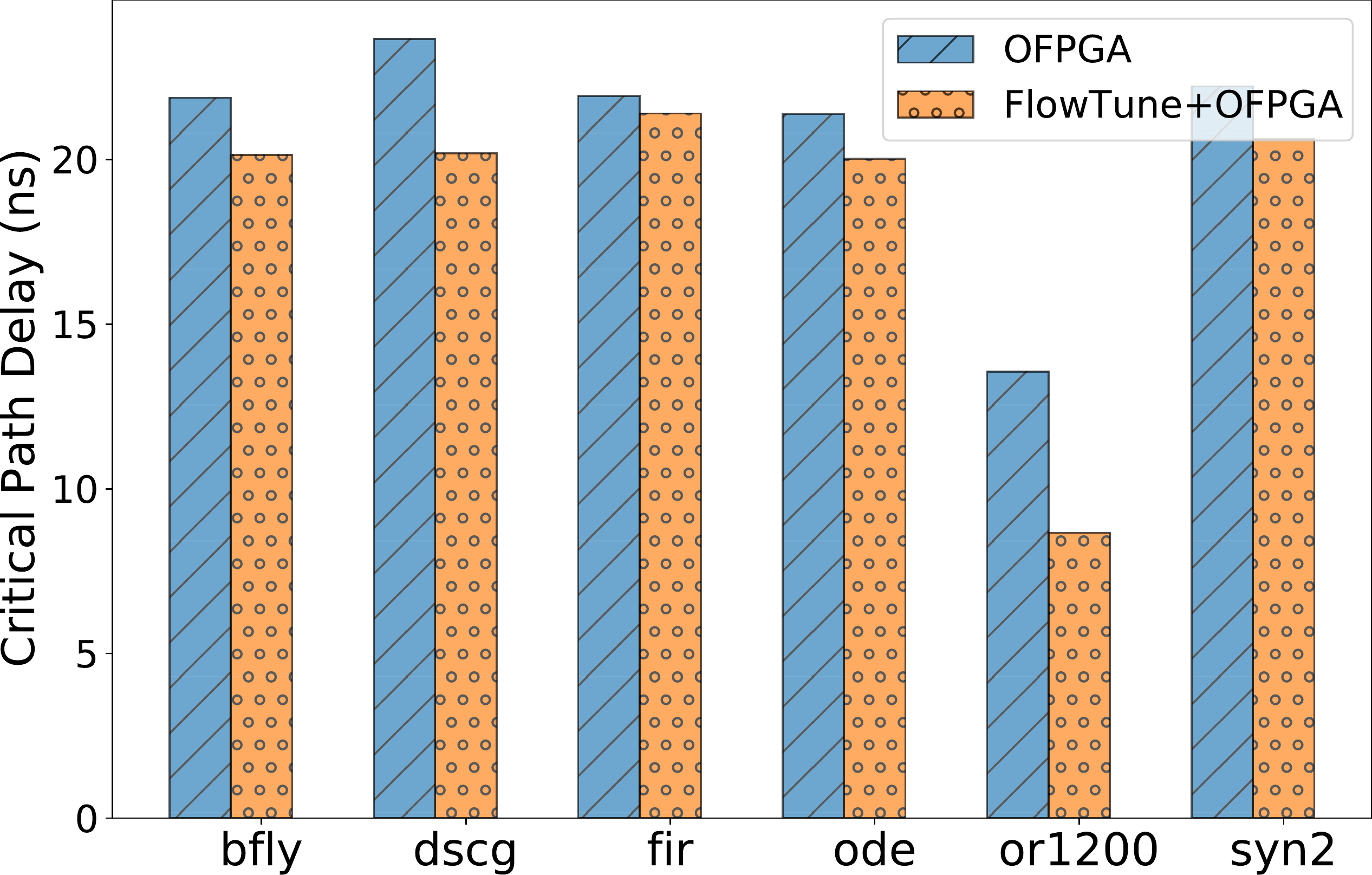}
\subcaption{ Post-PnR critical path delay.}\label{fig:ofpga_delay}
\end{minipage}
\caption{ Post-routing evaluation with LSOracle+OpenFPGA where the logic synthesis process involves \textbf{MIG logic optimization}, with OpenFPGA as backend using the same benchmarks collected from \cite{luu2014vtr}. The results are compared to default LSOracle+OpenFPGA flow, including (a) total area including logic and routing area, (b) total routing wire length, (c) post-PnR total negative slacks (TNS), and (d) post-PnR critical path delay.}
\label{fig:result_ofpga}
\end{figure*}

\medskip
\noindent
\textbf{Results on total area} -- (1) \textbf{VTR results} -- Figure \ref{fig:vpr_area} shows the results of post-routing evaluation on design area which is the sum of total used logic block area and the total routing area. When designs are evaluated with \textit{FlowTune}, the area can be reduced by $\sim30\%$ on average for \texttt{bfly}, \texttt{dscg}, \texttt{fir}, \texttt{ode}, and \texttt{syn2}. 
However, for design \texttt{or1200}, we can achieve few improvement results with \textit{FlowTune} optimization. (2) \textbf{OpenFPGA results} -- OpenFPGA results show area improvements in all the cases, as seen in Figure  \ref{fig:ofpga_area}. Results present up to 10.0\% area improvement, with an average of 4.5\% area reduction. While these results are not as expressive as in the VTR flow, it is still relevant and consistent among both flows. 

\medskip
\noindent
\textbf{Results on total wirelength} -- (1) \textbf{VTR results} -- We further analyze the total wirelength of the post-PnR designs to clearly show that routing has been improved, in addition to the logic optimization results. Figure \ref{fig:vpr_wl} shows the results of post-routing evaluation on total routing wirelength. The performance with \textit{FlowTune} can be improved for all designs. 
Similarly, for total area evaluation, the improvement is marginal for design \texttt{or1200}. (2) \textbf{OpenFPGA results} -- Figure  \ref{fig:ofpga_wl} presents the total wirelength for MIGs with OpenFPGA. In this case, \textit{FlowTune} presents better results in 4 cases, with little overhead in two designs (up to 4\%). On the other hand, \textit{FlowTune} reduces the total wirelength in up to 14\%, and 5\% on average. As the goal of \textit{FlowTune} is to reduce the logic area, it might reflect in longer total wirelength in some cases due to the improved logic sharing. Still, the overhead compared to the baseline flow was small for the considered designs. We can observe that \textit{FlowTune} was able to consistently reduce total wirelength in both flows for almost all the benchmarks. When considering \texttt{or1200}, in the VTR flow, \textit{FlowTune} had minor gains compared to the baseline, whereas in OpenFPGA it had some overhead compared to the baseline. 

\medskip
\noindent
\textbf{Results on timing} -- Post-PnR timing results are presented with critcal path delay and total negative slacks (TNS), which are the most critical two metrics used for timing evaluation. (1) \textbf{VTR results} -- The performance with \textit{FlowTune} can be improved for all designs except for \texttt{or1200}. 
On the other hand, we have observed that the \texttt{or1200} timing performance has a slight improvement for TNS but got worse than the default in critical path delay. While there has been very little logic reduction (see Figure \ref{fig:vpr_area}) and wirelength reductions, the structure of the \texttt{or1200} design remains almost the same. We believe that the critical path delay and TNS results differences between with and w/o \textit{FlowTune} is from the randomness of the placement and routing algorithms in VPR. 
(2) \textbf{OpenFPGA results} -- {Figure  \ref{fig:ofpga_delay} presents the post-PnR critical path delay for the considered benchmarks.} We can see that \textit{FlowTune} greatly improves the baseline. In particular, we improve all the cases, with up to 37\% gains for the \textit{or1200}. Still, five designs have over 7\% gains in the WNS, showing the effectiveness of \textit{FlowTune} in improving MIGs over a state-of-the-art recipe. On average, \textit{FlowTune} reduces the delay by 7\%. Therefore, \textit{FlowTune} achieves great delay improvements in both flows. The main difference is that while in VTR flow it could not benefit the \texttt{or1200}, in the OpenFPGA flow the \texttt{or1200} was greatly improved. Figure  \ref{fig:ofpga_tns} presents the OpenFPGA total negative slack results. We improved the baseline in 5 cases, up to 12\%. For one case, we had a 1\% TNS overhead. On average, we presented an average TNS reduction of ~5\%. Total negative slack has a similar trend in both, with gains when applying \textit{FlowTune}.

In conclusion, we can see that FlowTune framework with the domain-specific MAB algorithm can flexibly and effectively navigate flow optimizations in different logic synthesis DAG representations, and different PnR backends, and yet produce effective and consistent results. That positions \textit{FlowTune} as a versatile, light-weight, and portable flow exploration framework.

\section{Related Work}

Recently, we have seen significant progress in leveraging ML techniques for logic synthesis. Specifically for exploring synthesis flows as sequential decision making problem, there are mainly two directions -- (1) learning a static ML-based predictor to enable fast design space and (2) exploring flows in reinforcement and iterative fashion. 

In \cite{DBLP:conf/dac/YuXM18}, Yu et al. proposed the first ML-based flow exploration approach, which involves a CNN-based QoR predictor to enable fast flow exploration and generation. They model the problem as a multi-class classification problem, and the CNN outputs angel- and devil-flows, where angel flows produce the best QoR results and devil flows likely offers the worst QoR. While this approach learns a static ML model that eliminates the expensive runtime of evaluating a large number of flows in synthesis tool, one critical drawback is on the data collecting and labelling. Although, the authors proposed a follow-up approach based on recurrent neural networks and transfer learning to reduce the efforts in collecting labelled dataset, this approach is limited on limited technology domain, which is limited on generalizability for technology transferability \cite{yu2020decision}.

{To work around the challenge of labelled data collection, reinforcement learning (RL) approaches has then been adopted for logic synthesis flow generation}. Liu et. al. \cite{liu2017parallelized} first cast logic optimization as a \textit{Markov Decision Process} (MDP), where a set of logic transformations could be chosen in the next iteration of the synthesis flow. However, it has been demonstrated that the performance of a sequence of logic synthesis transformations does not satisfy MDP properties, since the performance of each transformation does not solely depend on previous transformation.  Hosny et al. \cite{hosny2019drills} propose a Deep RL-based approach that aims to optimize the area given a timing constraint. They cast the generation of logic synthesis flow into a game-like problem, where the actions are the set of transformations to be selected. However, the RL training process is very time consuming and offers poor flexibility (e.g., limited flow lenghth, QoR-specific RL model, etc.,) and integration capability. 






To further ehance the structure information in the ML-based approaches for flow exploration, there have recently seen many works leveraging Graph Neural Network (GNNs) to improve the generalizability. Zhu et. al \cite{panrl} propose to model the logic synthesis flow generation as MDP problem and use GNNs to enhance the state representation. Therefore, besides AIG statistics and the history of transforms applied, they also aggregate the graph-structure through GNNs. They present improvements over the ABC \textit{resyn2} flow, with the same length of transformations to be applied. Similarly, in \cite{huaweirl}, the authors combine RL and GNN to optimize MIGs. In addition to utilize the feature extraction capability of GNNs, Wu \cite{wu2022hybrid} et. al. demonstrates that GNNs can be used to aggregate structure features such that static ML approaches can be trained with significantly reduced labeled data. Unfortunately, these work are only evaluated at the stage of logic-level without considering the impacts of low-level design stages such as placement and routing. 

\section{Conclusion}

This work proposes a multi-stage multi-armed bandit framework for Boolean logic optimization that is general end-to-end and high-performance domain-specific. We present an MAB-based synthesis flow exploration technique that takes advantage of domain-specific knowledge of DAG-aware synthesis algorithms. To highlight the value of the collected domain information, a complete analysis of DAG-aware algorithms in synthesis flows is offered. We also present a novel MAB mechanism, which includes a rapid startup and a multi-stage MAB exploration strategy. We built a complete exploration framework that interfaces with numerous tools to illustrate the performance and versatility of our framework. Our results show that FlowTune outperforms prior works in terms of optimization efficiency and runtime for standard-cell technology mapping, as well as end-to-end PnR assessment using various backend tools. This is the first framework that shows end-to-end synthesis experiments in terms of post-PnR performance indicators. We also show that our domain-specific MAB algorithm can be applied to a variety of DAG-based logic synthesis, with FlowTune being used for both AIG and MIG improvements. Explainability and robustness analysis of ML-based design space exploration, as well as architecture-aware optimizations will be the focus of future work.

\textbf{Acknowledgement} This work is funded by National Science Foundation (NSF) NSF-2008144, DARPA IDEA, and NSF CAREER awards NSF-1751064 and NSF-2047176.

\bibliographystyle{IEEEtran}
\bibliography{synthesis}

%

\begin{IEEEbiography}[{\includegraphics[width=1in,height=1.25in,clip,keepaspectratio]{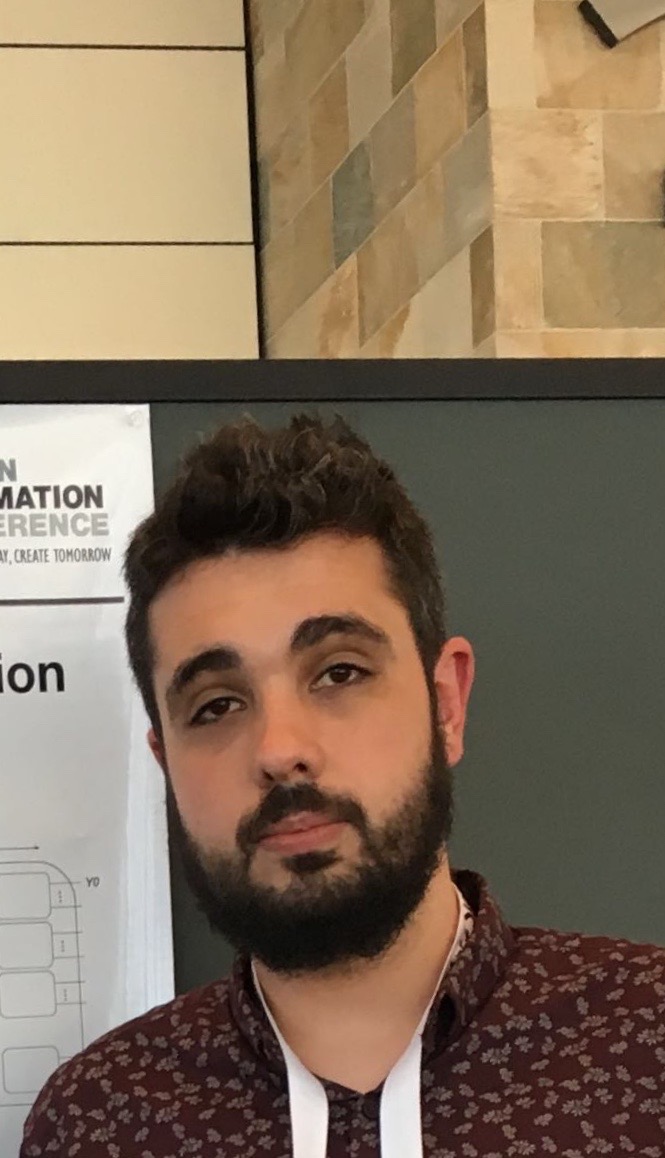}}]{Walter Lau Neto}
received the M.S. degree in Microelectronics from the Universidade Federal do Rio Grande do Sul (UFRGS), Porto Alegre, Brazil, in 2018. In 2022, he received the Ph.D. degree in Computer Engineering from the University of Utah, Salt Lake City, UT, USA. 
Currently, he is a Senior R\&D engineer at Synopsys Inc., Sunnyvale, CA, USA, where he works on developing novel logic synthesis techniques. He is a reviewer for several conferences and journals, and served in the Organizing Committee of IWLS'2021 and IWLS'2022. His research interests include logic synthesis, machine learning for logic synthesis, and electronic design automation. 

\end{IEEEbiography}
\begin{IEEEbiography}[{\includegraphics[width=1in,height=1.25in,clip,keepaspectratio]{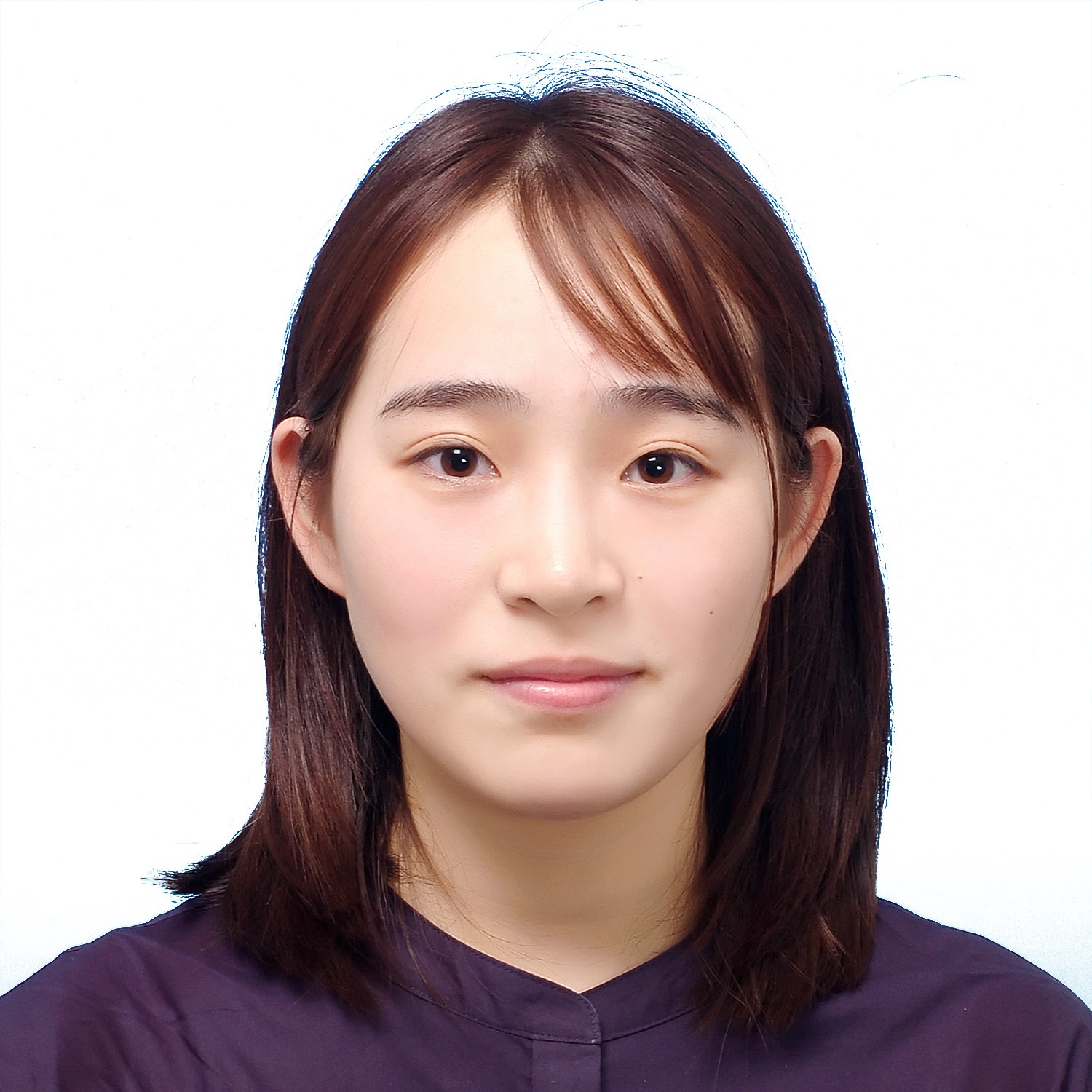}}]{Yingjie Li} (S'21) is pursuing Ph.D. at the University of Utah under the supervision of Prof. Cunxi Yu, starting from 2020 Spring. Her research interests lie in optical neural networks including building the compilation systems and developing new algorithms, and electronic design automation including logic synthesis and physical design. She received B.S. degree in 2018 from Huazhong University of Science and Technology in Wuhan, China, and her master degree from Cornell University in 2019. Her recent research interests focus on hardware-software codesign and compilation systems for optical neural networks and physics-aware adversarial machine learning, and machine learning for EDA. 

\end{IEEEbiography}
\begin{IEEEbiography}[{\includegraphics[width=1in,height=1.25in,clip,keepaspectratio]{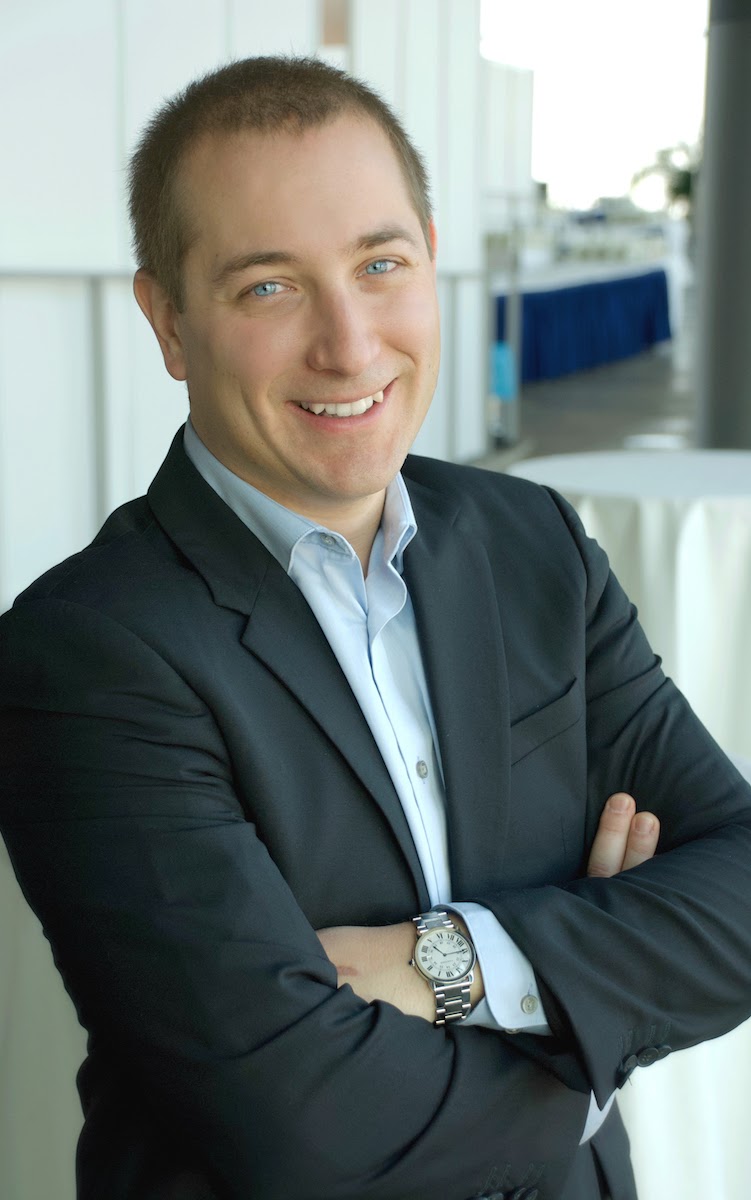}}]{Pierre-Emmanuel Gaillardon}(S’10–M’11–SM’16) is an Associate Professor and the Associate Chair for Academics and Strategic Initiatives in the Electrical and Computer Engineering (ECE) department and an adjunct Associate Professor in the School of Computing at The University of Utah, Salt Lake City, UT, where he leads the Laboratory for NanoIntegrated Systems (LNIS). He holds an Electrical Engineer M.Sc. degree from CPE-Lyon, France (2008), a M.Sc. degree in Electrical Engineering from INSA Lyon, France (2008) and a Ph.D. degree in Electrical Engineering from CEA-LETI, Grenoble, France and the University of Lyon, France (2011).

Prior to joining the University of Utah, he was a research associate at the Swiss Federal Institute of Technology (EPFL), Lausanne, Switzerland within the Laboratory of Integrated Systems (Prof. De Micheli) and a visiting research associate at Stanford University, Palo Alto, CA, USA. Previously, he was research assistant at CEA-LETI, Grenoble, France. Prof. Gaillardon is recipient of the C-Innov 2011 best thesis award, the Nanoarch 2012 best paper award, the BSF 2017 Prof. Pazy Memorial Research Award, the 2017 NSF CAREER award, the 2018 IEEE CEDA Pederson Award, the 2018 ChemE Education William H. Corcoran best paper award, the 2019 DARPA Young Faculty Award, the 2019 IEEE CEDA Ernest S. Kuh Early Career Award and the 2020 ACM SIGDA Outstanding New Faculty Award. He has been serving as TPC member for many conferences, including DATE, DAC, ICCAD, Nanoarch, etc.. He is an associate editor of IEEE TNANO and a reviewer for several journals and funding agencies. He served as Topic co-chair "Emerging Technologies for Future Memories" for DATE'17-19. He is a senior member of the IEEE.
\end{IEEEbiography}
\begin{IEEEbiography}[{\includegraphics[width=1in,height=1.25in,clip,keepaspectratio]{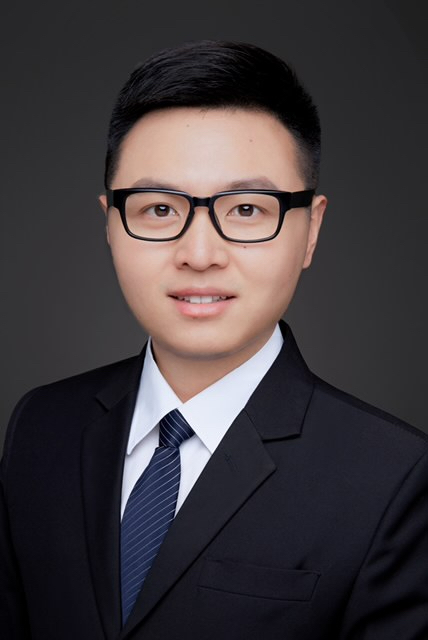}}]{Cunxi Yu} (S'15-M'17)
is an Assistant Professor in the ECE Department at the University of Utah. His research interests focus on novel algorithms, systems, and hardware designs for computing and security. Before joining University of Utah, Cunxi was a PostDoc at Cornell University in 2018-2019 and EPFL in 2017-2018, and was a research intern at IBM T.J Watson Research Center (2015, 2016). He received Ph.D. degree from UMass Amherst in 2017. His work received the best paper nomination at ASP-DAC (2017), TCAD Best paper nomination (2018), 1st place at DAC Security Contest (2017), NSF CAREER Award (2021), and DLS Best Poster Honorable Mention at (2022). He served as Organizing Committee in IWLS, ICCD, VLSI-SoC, ASAP, as a TPC member in ICCAD, DATE, ASP-DAC, DAC, and General Chair of IWLS 2023.
\end{IEEEbiography}





\vfill

\end{document}